\documentclass[useAMS,usenatbib]{mnras}
\usepackage{graphicx}

\usepackage{journal_names}
\usepackage{subfig}
\usepackage{pdflscape}
\usepackage{amssymb}

\title[What are the MPS sources?]{What are the megahertz peaked-spectrum sources?}
\author[Rocco Coppejans et al.]{Rocco Coppejans$^{1}$\thanks{E-mail: r.coppejans@astro.ru.nl }, D\'{a}vid Cseh$^{1}$, Sjoert van Velzen$^{2}$, Heino Falcke$^{1,3}$, \newauthor Huib T. Intema$^{4}$, Zsolt Paragi$^{5}$, Cornelia M\"{u}ller$^{1}$, Wendy L. Williams$^{6,4,3}$, \newauthor S\'{a}ndor Frey$^{7}$, Leonid I. Gurvits$^{5,8}$ and Elmar G. K\"{o}rding$^{1}$\\
$^{1}$Department of Astrophysics/IMAPP, Radboud University Nijmegen, PO Box 9010, 6500 GL Nijmegen, The Netherlands\\
$^{2}$Department of Physics and Astronomy, The Johns Hopkins University, Baltimore, MD 21218, USA\\
$^{3}$Netherlands Institute for Radio Astronomy (ASTRON), PO Box 2, 7990 AA, Dwingeloo, The Netherlands\\
$^{4}$Leiden Observatory, Leiden University, PO Box 9513, 2300 RA, Leiden, The Netherlands\\
$^{5}$Joint Institute for VLBI ERIC, Postbus 2, 7990 AA Dwingeloo, The Netherlands\\
$^{6}$School of Physics, Astronomy and Mathematics, University of Hertfordshire, College Lane, Hatfield AL10 9AB, UK\\
$^{7}$F\"{O}MI Satellite Geodetic Observatory, PO Box 585, H-1592 Budapest, Hungary\\
$^{8}$Department of Astrodynamics \& Space Missions, Delft University of Technology, 2629 HS Delft, Delft, The Netherlands}

\begin{document}

\date{}

\pagerange{\pageref{firstpage}--\pageref{lastpage}} \pubyear{2014}

\maketitle

\label{firstpage}

\begin{abstract}
Megahertz peaked-spectrum (MPS) sources have spectra that peak at frequencies below 1\,GHz in the observer's frame and are believed to be radio-loud active galactic nuclei (AGN). We recently presented a new method to search for high-redshift AGN by identifying unusually compact MPS sources. In this paper, we present European VLBI Network (EVN) observations of 11 MPS sources which we use to determine their sizes and investigate the nature of the sources with $\sim10$\,mas resolution. Of the 11 sources, we detect nine with the EVN. Combining the EVN observations with spectral and redshift information, we show that the detected sources are all AGN with linear sizes smaller than 1.1\,kpc and are likely young. This shows that low-frequency colour-colour diagrams are an easy and efficient way of selecting small AGN and explains our high detection fraction ($82$\,per\,cent) in comparison to comparable surveys. Finally we argue that the detected sources are all likely compact symmetric objects and that none of the sources are blazars.
\end{abstract}

\begin{keywords}
techniques: interferometric -- high angular resolution -- radio continuum: galaxies -- galaxies: active -- galaxies: high-redshift
\end{keywords}

\section{Introduction}
\label{sec:introduction}

AGN jets are powered by the accretion of material from their host galaxy onto a supermassive black hole \citep[e.g.][]{1979ApJ...232...34B,1995A&A...293..665F} and can grow to extend well beyond their host galaxy. The young jets can be distorted and even stopped by the ambient medium, while larger jets heat both the interstellar and intergalactic medium, quench star formation and expel material from the galaxy \citep[e.g.][]{2013Sci...341.1082M}. Hence to understand AGN, we need to understand galaxies and vice versa \citep[e.g.][]{fabian2012}.

Young or restarted AGN can be used to study how AGN are launched and evolve from parsec-scale objects to sources of hundreds of kiloparsec such as Cygnus A and 3C175 \citep[e.g.][]{snellen2000}. AGN also act as beacons allowing us to observe sources out to $z>7$ \citep[e.g.][]{mortlock2011} and study how the Universe evolved from a time when it was less than 6\,per\,cent of its current age. Specifically, the fraction of jets that are frustrated by their host galaxy appears to increase with redshift \citep{2015MNRAS.446.2985V}. By comparing the number of young and small AGN at high redshifts to those in the modern Universe, we can therefore trace the evolution of the ambient medium which both feeds AGN and hampers or even confines their jets \citep[e.g.][]{falcke2004}. Hence, searching for both young AGN and AGN at high redshifts is critically important to understand what triggers the nuclear activity, how AGN evolve in size, how the population evolves with redshift and, ultimately, the origin of their redshift evolution. 

Compact steep-spectrum (CSS), gigahertz peaked-spectrum (GPS) and high-frequency peaked (HFP) sources are all radio-loud AGN that are identified based on their spectral energy distribution in the radio. CSS, GPS and HFP sources are characterised by steep optically thin spectra that turn over and have inverted spectra (the spectral index, $\alpha$, is defined as $S \propto \nu^{\alpha}$, where $S$ is the flux density and $\nu$ is the frequency) above the turnover frequency. The CSS sources have typical rest-frame turnover frequencies ($\nu_{\mathrm r}$) smaller than $500$\,MHz and largest linear sizes (LLS) of $1-20$\,kpc \citep{o'dea1998}. For the GPS sources, $1<\nu_{\mathrm r}<5$\,GHz, while $\nu_{\mathrm r}>5$\,GHz for the HFP sources \citep{2000A&A...363..887D}. Both the GPS and HFP sources have $\mathrm{LLS}<1$\,kpc \citep{o'dea1998}. 

Morphologically GPS and HFP sources are typically classified as compact symmetric objects (CSOs) while the CSS sources are medium-size symmetric objects (MSOs) \citep[e.g.][]{snellen2000,Conway2002}. CSOs and MSOs are characterised by unbeamed emission from their steep-spectrum radio lobes on either side of a central position and have sizes smaller than their host galaxy \citep{Fanti1995, 2009AN....330..120F}. To strictly classify a source as a CSO or MSO, its flat-spectrum core has to be detected \citep{Orienti2014}, which is often not the case. In addition, unlike their names suggest, CSOs and MSOs are often not symmetric around their cores. This is likely the result of the interaction of the jet with an inhomogeneous ambient medium \citep[e.g.][]{o'dea1998,Orienti2014}.

Based on spectral and kinematic age estimates (determined by fitting models to the source spectra and measuring lobe expansion speeds), most of the HFP, GPS and CSS sources are believed to be young ($\lesssim 10^5$\,years) and small ($\lesssim20$\,kpc) AGN rather than being sources that are confined by the ambient medium of their host galaxy \citep{o'dea1998,Murgia2002,Conway2002,2003PASA...20...19M,2009AN....330..120F}. In addition, an empirical relation was found between $\nu_{\mathrm r}$ and LLS that spans three orders of magnitude \citep[see Section \ref{subsec:what are mps},][]{o'dea1998,snellen2000,Orienti2014}. This shows that the smallest sources have the highest turnover frequencies. Based on this evidence, it is believed that the HFP sources evolve into GPS sources which in turn evolve into CSS sources \citep{o'dea1998,snellen2000,tschager2003}. There is also strong evidence from their expected luminosity evolution and the similarities between their host galaxies  \citep{Begelman1996,snellen2000,devries2002} that the CSS sources will evolve into the FR\,I and FR\,II radio galaxies \citep{1974MNRAS.167P..31F}.

In the past, searches for high-redshift AGN in the radio have focused on ultra-steep-spectrum (USS) sources \citep[e.g.][]{jarvis2001,cruz2006,deBreuck2006}. However, the reason why USS sources should be at higher redshifts than non-USS sources remains unclear \citep{miley2008}. Moreover, several recent studies have found that USS sources are not at higher redshifts than non-USS sources \citep{ker2012,singh2014,smolcic2014}. 

In our previous paper \citep{coppejans2015}, we described a new method of searching for high-redshift radio-loud AGN by selecting compact megahertz peaked-spectrum (MPS; turnover frequency below 1\,GHz) sources. The MPS sources are believed to be a mixture of nearby CSS sources, and smaller GPS and HFP sources whose spectral turnovers have been redshifted to lower frequencies. Hence, the most compact MPS sources should be at the highest redshifts. In \citet{coppejans2015}, we took the first steps in testing the method by making a low-frequency radio colour-colour diagram and selecting a sample of 33 MPS sources from it. Using their photometric redshifts, we concluded that there is encouraging evidence that the MPS method can be used to search for high-redshift AGN. This was the first time that a colour-colour diagram was used to select MPS sources. However, it will soon be possible to repeat the analysis over the full sky using instruments such as the Low-Frequency Array \citep[LOFAR;][]{lofar}. We therefore wish to confirm that this novel selection method yields a separate class of small and likely young AGN, where the most compact sources are at high redshifts.

Here we present very long baseline interferometry (VLBI) observations of 11 MPS sources conducted with the European VLBI Network (EVN). Combining the EVN's sub-arcsecond resolution with the spectra of the sources, we investigate the nature of the MPS sources and test the hypothesis that these are AGN with small jets. In Section \ref{sec:data reduction}, we describe how we selected the sources and reduced the EVN data. Section \ref{sec: results and discussion} describes how the source spectra were generated, presents the source properties derived from the images and spectra, and discusses whether the radio emission is from star formation in the host galaxy or AGN activity. The individual sources are discussed in detail in Section \ref{sec:Comments on Individual Sources}. In Section \ref{sec:summary+conclude}, we discuss what the MPS sources are and present a summary in Section \ref{sec:Summary}. Throughout this paper, we use the following cosmological parameters: $\Omega_{\rm m}=0.3$, $\Omega_{\lambda}=0.7$, $H_0=72$\,km\,s$^{-1}$\,Mpc$^{-1}$.


\section[]{Target Selection, Observations and Data Reduction}
\label{sec:data reduction}

\subsection{Target Selection}
\label{subsec:Source Selection}
In \citet{coppejans2015}, we matched the sources in our 324.5\,MHz Very Large Array (VLA) P-band image (hereafter referred to as the VLA-P image) of the National Optical Astronomy Observatory (NOAO) deep wide-field survey Bo\"{o}tes field to the VLA Faint Images of the Radio Sky at Twenty-Centimeters (FIRST) survey \citep{first} and a 153\,MHz Giant Metrewave Radio Telescope (GMRT) catalogue of the field \cite[][hereafter referred to as the WIR catalogue]{Williams2013}. From this we generated a colour-colour diagram of the field and selected 33 MPS sources that either show a turnover in their spectra or a significant low-frequency flattening, which could indicate a turnover below 153\,MHz. Sources were excluded if they were extended in the FIRST or VLA-P catalogues or had a flux density difference of more than 20\,per\,cent between any two of the following three 1.4\,GHz catalogues: FIRST, National Radio Astronomy Observatory (NRAO) VLA Sky Survey \cite[NVSS;][]{nvss} and \citet[][hereafter referred to as the dVMR catalogue]{deVries2002bootes}. Since the resolution of the FIRST, dVMR and NVSS catalogues are 5.4, 20 and 45\,arcsec, respectively (see Table \ref{tbl:matching catalogues}), this not only removed variable sources, but also sources with extended structures that are resolved out in one of the higher resolution catalogues.

The MPS sources presented in this paper are given in Table~\ref{tbl:targets} and were observed with the EVN during two projects, EV020 and EC053. In the table the columns are: (1) source name, (2) the EVN project code under which the source was observed, (3) low-frequency spectral index calculated between 153 and 325\,MHz, (4) high-frequency spectral index calculated between 325 and 1400\,MHz, (5,6) photometric redshifts calculated using the \textsc{eazy} code from \citet{brammer2008} and \textsc{lrt} code from \citet{assef2008} as described in \citet{coppejans2015}, and (7) the total time spent observing the source with the EVN. A description of how $\alpha_{\mathrm{low}}$ and $\alpha_{\mathrm{high}}$ were calculated are given in Section \ref{subsec:Matched Catalogues}. Note that for the redshift values, the \textsc{lrt} code does not provide errors and includes an empirical AGN spectral energy distribution (SED) template in the fitting, and should therefore fit AGN spectra better. In the table, there are three sources (J142850+345420, J143718+364549 and J144230+355735) for which we do not have photometric redshifts. J142850+345420's optical counterpart is too faint for us to find a photometric redshift for it, while J143718+364549 and J144230+355735 lie outside the multiwavelength coverage of the Bo\"{o}tes field.

\begin{table*}
 \centering
 \begin{minipage}{12cm}
  \caption{Basic parameters of the target sources.}
  \begin{tabular}{ccccccc}
  \hline
  Source ID &  Project code & $\alpha_{\mathrm{low}}$ $^{\mathrm a}$ & $\alpha_{\mathrm{high}}$ $^{\mathrm b}$ & $z_{\rm \textsc{eazy}}$ & $z_{\rm \textsc{lrt}}$ & Time on\\
            &               &                &                 &                         &                        & source [min]\\
  (1)       &  (2)          & (3)            & (4)             & (5)                     & (6)                    & (7)      \\  
  \hline  
  J142850+345420 & EV020 & $0.6\pm0.4$  & $-0.5\pm0.1$ & ---                       & ---  & 45\\
  J142904+354425 & EC053 & $0.0\pm0.5$  & $-0.6\pm0.1$ & $0.809^{+0.084}_{-0.081}$ & 0.84 & 120\\  
  J142917+332626 & EC053 & $0.6\pm0.4$  & $-0.6\pm0.1$ & $1.583^{+0.322}_{-0.290}$ & 2.49 & 70\\  
  J143024+352438 & EV020 & $0.2\pm0.6$  & $-0.7\pm0.1$ & $1.196^{+0.116}_{-0.118}$ & 1.33 & 200\\
  J143042+351240 & EC053 & $0.0\pm0.4$  & $-0.7\pm0.1$ & $1.281^{+0.202}_{-0.217}$ & 1.13 & 200\\
  J143050+342614 & EV020 & $0.5\pm0.3$  & $-0.6\pm0.0$ & $2.364^{+0.535}_{-0.536}$ & 2.98 & 45\\
  J143055+350852 & EC053 & $-0.1\pm0.3$  & $-0.8\pm0.1$ & $2.195^{+0.379}_{-0.397}$ & 0.38 & 60\\  
  J143213+350940 & EV020 & $-0.1\pm0.3$  & $-0.3\pm0.1$ & $0.978^{+0.100}_{-0.095}$ & 0.96 & 45\\
  J143329+355042 & EC053 & $0.2\pm0.5$  & $-0.5\pm0.1$ & $2.821^{+2.032}_{-1.536}$ & 1.37 & 60\\
  J143718+364549 & EV020 & $0.2\pm0.5^{\mathrm c}$  & $-0.7\pm0.1^{\mathrm c}$ & ---                       & ---  & 45\\
  J144230+355735 & EV020 & $0.1\pm0.7^{\mathrm c}$  & $-1.0\pm0.2^{\mathrm c}$ & ---                       & ---  & 125\\
  \hline
  \multicolumn{7}{p{12cm}}{\footnotesize{\textbf{Notes:} $^{\mathrm a}$ $\alpha_{\mathrm{low}}$ was calculated between 153 and 325\,MHz. $^{\mathrm b}$ $\alpha_{\mathrm{high}}$ was calculated between 325 and 1400\,MHz. $^{\mathrm c}$ The spectral indices of the sources were calculated using their WENSS flux densities since they lie outside the area imaged with the VLA-P data (see Section \ref{subsec:Matched Catalogues}).}}\\
  \end{tabular}
  \label{tbl:targets}
 \end{minipage}
\end{table*}

The sources that were observed with the EVN were selected to be unresolved in FIRST, non-variable, have the highest possible flux density in FIRST and have not been previously observed with VLBI. Of the 11 MPS sources observed with the EVN, four are also in the selection of sources in \citet{coppejans2015}. Two of the new sources (J143718+364549 and J144230+355735) lie outside the region that was imaged with the VLA-P data and were selected based on their Westerbork Northern Sky Survey \cite[WENSS;][]{wenss} flux densities (see Section \ref{subsec:Matched Catalogues}). The remaining five new sources (J142850+345420, J143024+352438, J143042+351240, J143055+350852 and J143213+350940) were originally excluded in \citet{coppejans2015} because they have a flux density difference of more than 20\,per\,cent between either FIRST and dVMR or NVSS and dVMR. In Section \ref{subsec:variability}, we argue that the flux density difference with dVMR catalogue does not necessarily indicate that these sources are variable. We therefore believe that all five sources are genuine MPS sources.

\subsection{Observing Setup and Data Reduction}
\label{subsec:observing setup and Data Reduction}
EV020 and EC053 were observed on 15 April 2014 and 14 January 2015, respectively. Since $\alpha_{\mathrm{high}}<-0.5$ for the MPS sources, we elected to do the observations at 1.664\,GHz to maximize the flux density of the sources and reduce the required observing time. During both projects we requested the targets to be observed with the radio telescopes at Effelsberg (Germany), Jodrell Bank (Mk2; UK), Medicina (Italy), Noto (Italy), Onsala (Sweden), Toru\'{n} (Poland), Sheshan (China) and the Westerbork Synthesis Radio Telescope (WSRT, the Netherlands). A list of the telescopes and whether or not they successfully participated in each project is given in Table \ref{tbl:dishes}. Since the baselines to Sheshan (which did not take part in EC053) form the longest baselines, the typical restoring beam size of EC053 is $26\times32$\,mas compared to the $3\times10$\,mas of EV020. Both projects obtained data with 2\,s integrations at 1024\,Mbit\,$\mathrm{s^{-1}}$ in left and right circular polarizations with eight subbands per polarization and 16\,MHz of bandwidth per subband. The technique of electronic VLBI (e-VLBI) was used, where the data are not recorded at the telescopes but streamed to the central correlator using optical fiber networks in real time. The observations of the targets were interleaved with observations of two phase calibrators, J1430+3649 and J1422+3223. The phase solutions from J1422+3223 were used to correct J142917+332626, which is $1\fdg8$ away from J1422+3223. The solutions of J1430+3649 were used to correct the remaining sources which are separated from it by between $0\fdg5$ and $2\fdg5$.

\begin{table}
 \centering
 \begin{minipage}{12cm}
  \caption{Telescope participation in each project.}
  \begin{tabular}{ccc}
  \hline
  Radio dish &  EC053$^{\mathrm a}$ & EV020$^{\mathrm a}$ \\
  \hline  
  Effelsberg   & Yes & Yes \\
  Jodrell Bank & Yes & Yes \\
  Medicina     & No  & No  \\
  Noto         & No  & Yes \\
  Onsala       & Yes & Yes \\
  Toru\'{n}    & Yes & Yes \\
  Sheshan      & No  & Yes \\
  WSRT         & Yes & Yes \\
  \hline
  \multicolumn{3}{p{5cm}}{\footnotesize{\textbf{Notes:} $^{\mathrm a}$ ``Yes'' indicates that the telescope provided useful data while ``No'' indicates that it did not. }}\\
  \end{tabular}
  \label{tbl:dishes}
 \end{minipage}
\end{table}

The data were reduced using the \textsc{aips} \citep{aips} software package by calibrating the visibility amplitudes using antenna gains and system temperatures measured at the telescopes. Next fringe-fitting was performed on the two phase calibrators. The phase calibrators were then imaged in the Caltech \textsc{difmap} package \citep{difmap} by doing several iterations of CLEAN and phase self-calibration. A final round of amplitude self-calibration was done on the phase calibrators in \textsc{difmap} to determine the global antenna gain correction factors. The gain correction factors varied between one and five per\,cent and were applied to the visibility amplitudes of all the sources in \textsc{aips}. Using the clean component models derived for the phase calibrators in \textsc{difmap}, improved phase solutions were calculated for the phase calibrators in \textsc{aips}. These solutions were applied to the target sources before they were exported from \textsc{aips} for flagging and imaging in \textsc{difmap}. To check that we did not miss any source components, or that any of the sources were significantly offset from the phase center, we started off by making images that were at least $5\times5$\,arcsec in size. We then cleaned the identified components in smaller images using uniform weighting to get the best possible position accuracy for the components before switching to natural weighting. Since the target sources have flux densities of only a few mJy, we did not self-calibrate them. We finally imaged all of the sources using a uv-taper with a Gaussian value of 0.1 and a Gaussian radius of 15\,million\,wavelengths (M$\lambda$). The uv-taper downweights the visibilities on baselines to Sheshan, where the uv-plane is sampled the least. This decreases both the resolution and noise of the image, allowing for the detection of diffuse emission around the source.


\section{Results and General Discussion}
\label{sec: results and discussion}
In this section we will first discuss the catalogues to which the sources were matched, before presenting the results derived from the EVN observations and the spectra of the sources. This will be followed by a discussion of the 1.4\,GHz variability of the sources and the cause of the radio emission. 

\subsection{Matched Catalogues}
\label{subsec:Matched Catalogues}
Table \ref{tbl:matching catalogues} contains a list of all the radio catalogues and images to which the sources were matched to constrain their radio spectra. The source matching was done using the method described in Section 3.2 of \citet{coppejans2015}. In the final column of Table \ref{tbl:matching catalogues}, a list of the sources with which a positive match were found is given for each of the catalogues.

\begin{landscape}
\thispagestyle{empty}
\begin{table}
  \caption{Catalogues that were matched to the target sources.}
  \begin{tabular}{ccccc}
  \hline
  Catalogue & Frequency & Resolution & Image noise $^{\mathrm a}$ & Positive matches\\
  name      & [MHz]     & [$''$]     & [mJy\,beam$^{-1}$]   & \\
  \hline  
  FIRST     & 1400 & $5.4\times5.4$ & 0.15  & All\\
  NVSS      & 1400 & $45\times45$   & 0.45  & All except J142917+332626\\
  dVMR      & 1380 & $13\times27$   & 0.03  & J142850+345420, J143024+352438, J143042+351240, J143050+342614, J143055+350852 and J143213+350940\\
  GMRT-608  &  608 & $5.0\times5.0$ & 0.04  & J142850+345420, J142904+354425, J143024+352438, J143042+351240, J143050+342614 and J143055+350852\\
  VLA-P     &  325 & $5.1\times5.6$ & 0.2*  & All except J143718+364549 and J144230+355735\\
  WENSS     &  325 & $54\times54$   & 3.6   & J143050+342614, J143055+350852, J143213+350940, J143718+364549 and J144230+355735\\
  WIR       &  153 & $25\times25$   & 2.0*  & All\\
  LOFAR-150 &  150 & $5.6\times7.4$ & 0.11* & All except J143718+364549 and J144230+355735\\
  LOFAR-62  &   62 & $19\times31$   & 4.8   & None\\
  \hline
  \multicolumn{5}{p{20cm}}{\footnotesize{\textbf{Notes:} $^{\mathrm a}$ Typical catalogue noise values are quoted except for those marked with *, where the noise is measured at the center of the image.}}\\
  \end{tabular}
  \label{tbl:matching catalogues}
\end{table}
\end{landscape}

The GMRT-608 image referenced in Table \ref{tbl:matching catalogues} is a mosaic of a part of the Bo\"{o}tes field (at 608\,MHz). A mosaic of the entire field will be published once the observations have been completed. The image was made from GMRT observations of a part of the Bo\"otes field (project code 28\_064). Four pointings covering 1.95\,deg$^2$ were observed on 24 and 26 July 2015. Raw visibilities were recorded every eight seconds in two polarizations (RR and LL), using 512 frequency channels to cover 32.0\,MHz of bandwidth centred on 608\,MHz. The on-target time for each pointing was between 100 and 110\,minutes. Primary flux density calibration was done with 3C286 using the wide-band low-frequency flux density standard of \citet{Scaife2012}. The data reduction follows that of \citet{2014MNRAS.444.3130D} and \citet{2014MNRAS.444L..44B} and was done in three stages: (i) initial gain and bandpass calibration, (ii)  self-calibration, and (iii) direction-dependent ionospheric phase calibration using the software package \textsc{SPAM} \citep{2009A&A...501.1185I}. The combined final mosaic reaches a root mean square (rms) noise level of $\approx 40-70$\,$\mu$Jy\,beam$^{-1}$ with a resolution of $5\times 5$\,arcsec. We have checked the consistency of the flux density scale by interpolating between the WIR and dVMR catalogues.

The 150\,MHz LOFAR-150 catalogue was made from a new LOFAR survey of the Bo\"{o}tes field (Williams et al., submitted). The image has a resolution of $\sim6$\,arcsec with half of the $19\mathrm{\,deg^2}$ image having a local rms noise below $0.18\mathrm{\,mJy\,beam^{-1}}$, both of which are better than the WIR image. The catalogue itself contains 5652 sources detected above a threshold of $5\sigma$ (Williams et al., submitted). After matching our sources to the catalogue we found that all but one of the LOFAR-150 flux densities were higher than the corresponding WIR flux densities, despite the LOFAR-150 catalogue having a higher resolution. Specifically, the integrated LOFAR-150 flux densities for our sources have a median difference of 27\,per\,cent compared to those of the WIR catalogue. Since the LOFAR-150 flux density scale was checked and corrected using the flux densities of the sources in the WIR catalogue (Williams et al., submitted), we elected to use the WIR flux densities when fitting the source spectra (Section \ref{subsec:results}). We do however discuss (Section \ref{sec:Comments on Individual Sources}) and show (Fig. \ref{fig:spectra}) the LOFAR-150 flux densities for each of the sources in their spectral plots. We finally note that J143718+364549 and J144230+355735 fall outside the area imaged by the LOFAR-150 catalogue.

Finally, the LOFAR-62 catalogue \citep{vanweeren2014} is a catalogue constructed from LOFAR Low Band Antenna (LBA) commissioning observations at 62\,MHz of the Bo\"{o}tes field. While all our sources lie in the image, none of them are detected at the catalogue's $5\sigma$ detection threshold. To determine the detection threshold for each source (given in Table \ref{tbl:lofar-62_detect}), we measured the the local rms noise in a $320\times320$\,arcsec box, as was done by \citet{vanweeren2014}, centred on each of the sources' positions and multiplied it by five. We used the local noise, rather than the typical noise of the catalogue (given in Table \ref{tbl:matching catalogues}), since the noise at the position of each of our sources will likely differ from the typical noise. This helped us to constrain the spectra of some of the sources (see Fig. \ref{fig:spectra} and Section \ref{sec:Comments on Individual Sources}).

\begin{table}
 \centering
 \begin{minipage}{5cm}
  \caption{62MHz detection threshold for each source in the LOFAR-62 catalogue.}
  \begin{tabular}{cc}
  \hline
  Source ID &  Detection threshold \\
            &        [mJy]       \\
  \hline  
  J142850+345420 & 28.7\\
  J142904+354425 & 43.8\\  
  J142917+332626 & 75.3\\  
  J143024+352438 & 46.8\\
  J143042+351240 & 47.2\\
  J143050+342614 & 36.2\\
  J143055+350852 & 30.4\\  
  J143213+350940 & 32.2\\
  J143329+355042 & 36.1\\
  J143718+364549 & 50.9\\
  J144230+355735 & 68.1\\
  \hline
  \end{tabular}
  \label{tbl:lofar-62_detect}
 \end{minipage}
\end{table}

\begin{figure}
  \includegraphics[width=\columnwidth]{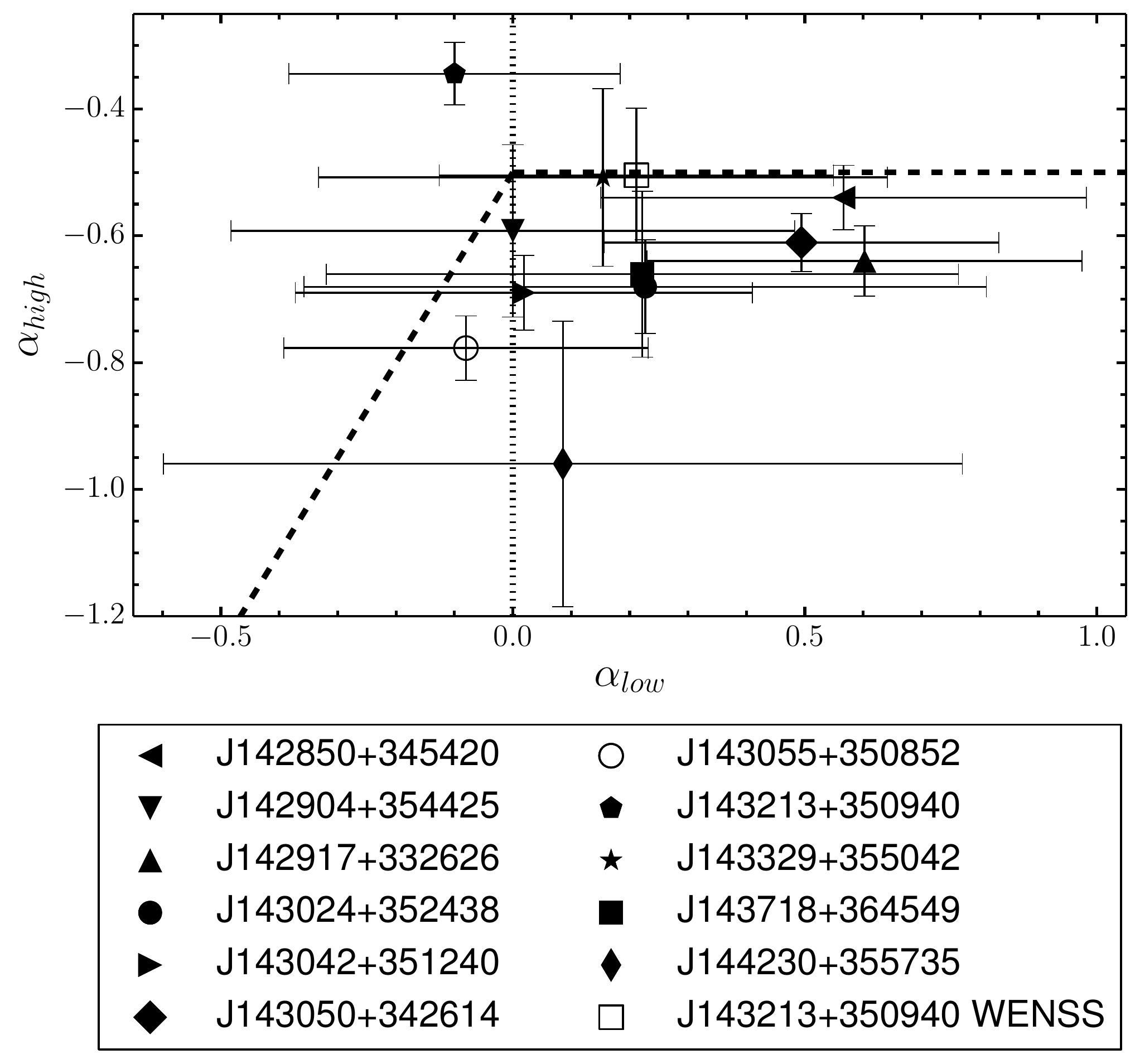}
  \caption{A colour-colour diagram of the sources. In the plot, $\alpha_{\mathrm{low}}$ and $\alpha_{\mathrm{high}}$ were calculated as explained in Section \ref{subsec:Matched Catalogues}. The dotted vertical line indicates a spectral index of zero while the MPS sources were selected from the region below and to the right of the dashed line.}
  \label{fig:colour-colour}
\end{figure}

In Fig. \ref{fig:colour-colour}, the colour-colour diagram for the sources is shown. For all the sources, except J143718+364549, J144230+355735 and J143213+350940 WENSS, $\alpha_{\mathrm{low}}$ was calculated between the flux densities of the WIR and VLA-P catalogues. $\alpha_{\mathrm{high}}$ was calculated between those of the VLA-P and FIRST catalogues. Since J143718+364549 and J144230+355735 are not in the VLA-P image, $\alpha_{\mathrm{low}}$ and $\alpha_{\mathrm{high}}$ were calculated from their WENSS flux densities. Since the VLA-P and WENSS flux densities differ significantly for J143213+350940, we also plotted J143213+350940's position in Fig. \ref{fig:colour-colour} if $\alpha_{\mathrm{low}}$ and $\alpha_{\mathrm{high}}$ are calculated using its WENSS flux density rather than the VLA-P flux density. This will be discussed in detail in Section \ref{subsec:J143213+350940}. The MPS sources in \citet{coppejans2015} were selected from the area below and to the right of the dotted lines in Fig. \ref{fig:colour-colour}, which are defined by $\alpha_{\mathrm{high}}<-0.5$ and $\alpha_{\mathrm{high}}<1.5\alpha_{\mathrm{low}} - 0.5$. The first constraint ensures that the sources have a steep spectrum above 1\,GHz. The second allows us to not only select the sources with a clear spectral peak, but also sources whose spectra flatten towards lower frequencies, which could indicate a spectral turnover below 153\,MHz. We note that in Fig. \ref{fig:colour-colour}, J143213+350940 does not satisfy the selection criteria, while J143213+350940 WENSS does (Section \ref{subsec:J143213+350940}).

\subsection{Source properties}
\label{subsec:results}
In Table \ref{tbl:derived_info}, the parameters derived for the sources are presented with the sources components named as the source name followed by a letter. These components are shown in the EVN images presented in Figures \ref{fig:J142850+345420} through \ref{fig:J144230+355735}.

\begin{table*}
 \centering
 \begin{minipage}{\textwidth}
  \caption{Derived parameters of the sources.}
  \begin{tabular}{ccccccc}
  \hline
  Source ID & Noise & RA $^{\mathrm g}$ & DEC $^{\mathrm a}$ & $S_i$ $^{\mathrm b}$ & Min & Maj \\
            & [$\mu \mathrm{Jy\,beam}^{-1}$] & [J2000] & [J2000] & [mJy] & [mas] & [mas] \\
  (1)       & (2) & (3) & (4) & (5) & (6) & (7) \\
  \hline  
  J142850+345420  & 250 & --- & --- & $8.16\pm0.77$  & --- & $47.6\pm0.7$ $^{\mathrm c}$ \\
  J142850+345420a & 250 & 14:28:50.46588(0.00007) & 34:54:20.8346(0.0011) & $6.25\pm0.61$ & $29.9\pm1.7$ & $29.9\pm1.7$ \\ 
  J142850+345420b & 250 & 14:28:50.46894(0.00015) & 34:54:20.8206(0.0022) & $1.91\pm0.47$ & $21.9\pm1.2$ & $21.9\pm1.2$ \\ 
  
  \hline
  J142904+354425 & 23  & 14:29:04.6$^{\mathrm *}$ & 35:44:25.1$^{\mathrm *}$ & ---  & --- & ---  \\  
  
  \hline
  J142917+332626 & 170 & 14:29:17.42200(0.00003) & 33:26:26.6001(0.0004) & $3.47\pm0.32$ & $13.0\pm0.5$ & $13.0\pm0.5$ \\  
  
  \hline
  J143024+352438 & 11 & 14:30:24.3$^{\mathrm *}$ & 35:24.38.1$^{\mathrm *}$ & --- & --- & ---  \\
  
  \hline
  J143042+351240 & 18 & 14:30:42.57796(0.00004) & 35:12:40.6253(0.0006) & $0.20\pm0.03$ & $12.1\pm1.5$ & $12.1\pm1.5$ \\ 
  
  \hline
  J143050+342614 & 198 & 14:30:50.90925(0.00003) & 34:26:14.1890(0.0004) & $3.53\pm0.50$ & $3.0\pm0.4$ & $15.4\pm1.8$  \\ 
  
  \hline
  J143055+350852  & 88  & --- & --- & $5.81\pm0.37$ & --- & $137.9\pm4.4$ $^{\mathrm c}$ \\ 
  J143055+350852a & 88  & 14:30:55.07509(0.00001) & 35:08:52.8445(0.0002) & $3.78\pm0.23$ & $13.1\pm0.4$ & $13.1\pm0.4$ \\ 
  J143055+350852b & 88  & 14:30:55.07953(0.00020) & 35:08:52.9619(0.0030) & $2.03\pm0.29$ & $47.0\pm5.9$ & $47.0\pm5.9$ \\ 
  
  \hline
  J143213+350940$^{\mathrm **}$  & 400 & --- & --- & $14.41\pm1.64$ & $30.1\pm2.2$ & $69.6\pm5.2$ \\  
  J143213+350940a$^{\mathrm **}$ & 201 & 14:32:13.54889(0.00016) & 35:09:40.8707(0.0023) & $5.35\pm0.86$ & $31.5\pm4.7$ & $31.5\pm4.7$ \\ 
  J143213+350940b$^{\mathrm **}$ & 201 & 14:32:13.55036(0.00001) & 35:09:40.8569(0.0002) & $2.81\pm0.32$ & $4.1\pm0.3$  & $4.1\pm0.3$ \\ 
  
  \hline
  J143329+355042 & 95 & 14:33:29.85779(0.00001) & 35:50:42.2509(0.0001) & $5.85\pm0.33$ & $4.0\pm0.1$ & $19.3\pm0.4$ \\
  
  \hline
  J143718+364549 & 175 & 14:37:18.09615(0.00009) & 36:45:49.8219(0.0013) & $6.07\pm0.53$ & $40.2\pm2.7$ & $40.2\pm2.7$ \\  
  
  \hline
  J144230+355735$^{\mathrm **}$  & 30 & --- & --- & $2.32\pm0.17$ & --- & $134.4\pm0.7$ $^{\mathrm c}$ \\ 
  J144230+355735a$^{\mathrm **}$ & 36 & 14:42:30.69966(0.00036) & 35:57:35.0093(0.0054) & $0.56\pm0.21$  & $29.4\pm10.8$ & $29.4\pm10.8$ \\ 
  J144230+355735b$^{\mathrm **}$ & 36 & 14:42:30.69602(0.00019) & 35:57:35.0979(0.0028) & $0.63\pm0.19$  & $11.4\pm3.3$ & $33.3\pm9.6$   \\ 
  J144230+355735c$^{\mathrm **}$ & 36 & 14:42:30.69416(0.00053) & 35:57:35.0776(0.0080) & $0.31\pm0.19$  & $26.3\pm15.8$ & $26.3\pm15.8$ \\ 
  
  \hline
  \multicolumn{7}{p{17cm}}{\footnotesize{\textbf{Notes:} $^{\mathrm a}$ The uncertainty, in arcseconds, is given in brackets after the value. $^{\mathrm b}$ For all the multi-component sources except J143213+350940 and J144230+355735 (see note $^{\mathrm **}$), the value is the sum of the integrated flux densities of their components. $^{\mathrm c}$ The value is the source size (distance between the centroids of the two furthest components). $^{\mathrm d}$ For sources with two values, the first was calculated using $z_{\mathrm{\textsc{eazy}}}$ and the second using $z_{\mathrm{\textsc{lrt}}}$. $^{\mathrm e}$ For sources without redshifts, the values were calculated using a redshift of zero. $^{\mathrm f}$ For sources with upper limits, the limit was calculated using $z=1$. $^{\mathrm g}$ The uncertainty, in seconds, is given in brackets after the value. $^{\mathrm *}$ Since the source was not detected with the EVN, the RA and DEC values are taken from FIRST. $^{\mathrm **}$ The values of the source were derived from the uv-tapered image while the values of the components were derived from the non-uv-tapered image (see Sections \ref{subsec:J143213+350940} and \ref{subsec:J144230+355735}).}}\\
  \end{tabular}
  \label{tbl:derived_info}
 \end{minipage}
\end{table*}

\begin{table*}
 \centering
 \begin{minipage}{\textwidth}
  \contcaption{}
  \begin{tabular}{cccccc}
  \hline
  Source ID & \% flux & $T_{\mathrm b}$ $^{\mathrm d,e}$ & $\nu_{\mathrm o}$ & $\nu_{\mathrm r}$ $^{\mathrm d}$ & LLS $^{\mathrm d,f}$ \\
            & density  & [$\times10^6$\,K] & [GHz] & [GHz] & [pc]\\
  (1)       & (8) & (9) & (10) & (11) & (12) \\
  \hline  
  J142850+345420  & $99\pm11$  & --- & $0.38\pm0.08$  & --- & $<370\pm17$\\
  J142850+345420a & --- & $>3.1\pm0.4$ & --- & --- & ---\\ 
  J142850+345420b & --- & $>1.8\pm0.5$ & --- & --- & ---\\ 
  
  \hline
  J142904+354425 & --- & --- & $0.22\pm0.13$  & $0.38\pm0.24$  \& $0.40\pm0.25$  & ---\\  
  
  \hline
  J142917+332626 & $57\pm6$ & $23.4\pm3.9$ \& $31.6\pm3.4$ & $0.39\pm0.06$ & $1.00\pm0.20$ \& $1.35\pm0.21$ & $107\pm8$ \& $102\pm7$\\  
  
  \hline
  J143024+352438 & --- & --- & $0.29\pm0.14$  & $0.63\pm0.32$ \& $0.67\pm0.33$  & --- \\
  
  \hline
  J143042+351240 & $4\pm1$ & $1.4\pm0.4$ \& $1.3\pm0.3$ & $0.23\pm0.11$ & $0.51\pm0.25$ \& $0.49\pm0.23$ & $97\pm14$ \& $96\pm12$\\ 
  
  \hline
  J143050+342614 & $33\pm5$ & $113.0\pm30.7$ \& $133.7\pm29.4$ & $0.34\pm0.08$ & $1.13\pm0.31$ \& $1.33\pm0.30$ & $122\pm20$ \& $115\pm15$ \\ 
  
  \hline
  J143055+350852  & $71\pm6$ & --- & $0.18\pm0.09$ & $0.56\pm0.30$ \& $0.24\pm0.13$ & $1105\pm99$ \& $698\pm56$\\ 
  J143055+350852a & --- & $31.5\pm4.6$ \& $13.6\pm1.0$ & --- & --- & ---\\ 
  J143055+350852b & --- & $1.3\pm0.3$ \& $0.6\pm0.1$ & --- & --- & ---\\ 
  
  \hline
  J143213+350940$^{\mathrm **}$  & $94\pm12$ & $6.1\pm1.0$ \& $6.0\pm0.9$ & $0.27\pm0.13$ & $0.54\pm0.26$ \& $0.53\pm0.25$ & $539\pm66$ \& $537\pm53$\\  
  J143213+350940a$^{\mathrm **}$ & --- & $4.7\pm1.3$ \& $4.7\pm1.2$ & --- & --- & ---\\ 
  J143213+350940b$^{\mathrm **}$ & --- & $145.7\pm24.0$ \& $144.3\pm22.6$ & --- & --- & ---\\ 
  
  \hline
  J143329+355042 & $92\pm9$ & $128.1\pm68.6$ \& $79.5\pm5.0$ & $0.35\pm0.10$ & $1.33\pm0.80$ \& $0.83\pm0.23$ & $139\pm26$ \& $158\pm8$\\
  
  \hline
  J143718+364549 & $68\pm9$ & $>1.7\pm0.2$ & $0.34\pm0.10$ & --- & $<313\pm29$\\  
  
  \hline
  J144230+355735$^{\mathrm **}$  & $69\pm14$ & --- & $0.30\pm0.10$ & --- & $<1046\pm28$ \\ 
  J144230+355735a$^{\mathrm **}$ & --- & $>0.3\pm0.2$ & --- & --- & ---\\ 
  J144230+355735b$^{\mathrm **}$ & --- & $>0.7\pm0.4$  & --- & --- & ---\\ 
  J144230+355735c$^{\mathrm **}$ & --- & $>0.2\pm0.2$  & --- & --- & ---\\ 
  
  \hline
  \end{tabular}
  \label{tbl:derived_info_cont}
 \end{minipage}
\end{table*}

Columns (2), (3) and (4) in Table \ref{tbl:derived_info} are the rms noise, the right ascension (RA) and declination (DEC), respectively, of each of the components of the source in the EVN image. The uncertainty of the RA and DEC are given in brackets after the value. The uncertainties were calculated using the equation given in \citet{chap14whitebible}, to this we added the uncertainty of the position of the phase calibrator from the VLBA calibrator list\footnote{http://www.vlba.nrao.edu/astro/calib/} (0.14\,mas and 0.11\,mas for J1422+3223 and J1430+3649, respectively), in quadrature. 

Column (5) gives the EVN integrated flux density at 1.7\,GHz. The values were determined by fitting circular Gaussian brightness distribution models in \textsc{difmap} to all of the sources and source components except J143050+342614, J143213+350940, J143329+355042 and J144230+355735b. These sources were fitted with elliptical Gaussians since the fit did not converge when fitting circular Gaussians or the circular fit clearly does not describe the flux density distribution of the source. Since \textsc{difmap} does not report an error on the integrated flux density, the errors were calculated using the equations in \citet{chap14whitebible} and adding an additional five per\,cent to account for the VLBI amplitude calibration uncertainty, as done by e.g. \citet{Frey2015} and \citet{An2012}. The integrated flux densities for all the multi-component sources except J143213+350940 and J144230+355735 are the sum of the individual components where the flux density errors were calculated by adding the errors of the individual components in quadrature. See Sections \ref{subsec:J143213+350940} and \ref{subsec:J144230+355735} on how the values for J143213+350940 and J144230+355735 were calculated.

Columns (6) and (7) contains the minor- and major-axis full width at half-maximum (FWHM) of the Gaussians fitted to the sources. The errors were calculated using the equations in \citet{chap14whitebible}. For the sources which were fitted with a circular Gaussian, the values in columns (6) and (7) are the same. For all of the sources, the values in column (7) were used as the source size. If the source was resolved into multiple components, the size was determined by calculating the distance between the centers of the two components that are the furthest apart, taking into account the uncertainties of the central positions. Note, however, that while J143213+350940 is composed of multiple components, we fitted both components simultaneously with a single elliptical Gaussian (see Section \ref{subsec:J143213+350940}). Hence the values reported for J143213+350940 are the minor- and major-axis of the fitted Gaussian.

Column (8) gives the percentage of the predicted flux density that was recovered from the image. The value was calculated using $100S_{\mathrm i}/S_{\mathrm{predicted}}$, where $S_{\mathrm i}$ is the integrated EVN flux density of the sources given in column (5) and $S_{\mathrm{predicted}}$ is the sources' predicted flux density at 1.7\,GHz. $S_{\mathrm{predicted}}$ was calculated using $\alpha_{\mathrm{high}}$ from Table \ref{tbl:targets} in combination with the equation $S_{\mathrm{predicted}}=k\nu^{\alpha}$. The constant $k$ was calculated for each source using its integrated FIRST flux density. The errors of the values in Column (8) were calculated by propagating the errors of $S_{\mathrm i}$ and $S_{\mathrm{predicted}}$.

The redshift-corrected brightness temperatures of the sources in column (9) were calculated using
\begin{equation}
 T_{\mathrm b} = 1.22\times10^{12}(1+z)\frac{S_{\mathrm i}}{\theta_{1}\theta_{2}\nu^2}
 \label{eq:Tb}
\end{equation}
\citep{Condon1982}. Here, $z$ is the redshift, $S_{\mathrm i}$ is the integrated flux density in Jy, $\theta_{1}$ and $\theta_{2}$ are the major- and minor-axis of the Gaussian fitted to the source in mas, and $\nu$ is the observing frequency in GHz. If the source component was fitted with a circular Gaussian, $\theta_{1}=\theta_{2}$. Since we have two photometric redshifts for each source, we opted to calculate two values for each source for the relevant parameters in Table \ref{tbl:derived_info}. Since the upper and lower uncertainties of $z_{\mathrm{\textsc{eazy}}}$ are not symmetrical, we used the larger of the two as the uncertainty to calculate the errors reported for the relevant parameters. For the \textsc{lrt} code, which does not report an uncertainty on the redshift, we used an uncertainty of zero. Finally, to get robust lower limits for $T_{\mathrm b}$ for the sources without redshifts, we used a redshift of zero for these sources.

Column (10) contains the fitted observed turnover frequency ($\nu_{\mathrm o}$) for each source. Following \citet{Orienti2007}, \citet{Scaife2012} and \citet{Orienti2014}, we calculate $\nu_{\mathrm o}$ by fitting a second order polynomial of the form $\log_{10}(S_{\mathrm i})= a(\log_{10}(\nu)-\log_{10}(\nu_{\mathrm o}))^2 + b$ to the spectral plot of each of the sources where $a$ and $b$ are constants. The spectral points to which that function was fitted are composed of the flux densities from the FIRST, GMRT-608, VLA-P and WIR catalogues and are shown in Fig. \ref{fig:spectra}. For J143718+364549 and J144230+355735, which do not have VLA-P flux densities, we used the WENSS flux density. Since this involves fitting a function with three unknown parameters to three data points for the sources without GMRT-608 flux densities, the error values reported by the fitting algorithm can not be trusted for these sources. To improve the error estimates, we used a Monte Carlo method to estimate $\nu_{\mathrm o}$ and its error for all the sources. To do this, we used a random number generator to find new flux density values at each frequency for the source and calculated a new value of $\nu_{\mathrm o}$. The flux densities returned by the random number generator are Gaussian distributed values centred on the original flux density with a standard deviation equal to the error on the flux density. Repeating the procedure 100,000 times, we generated a histogram of all the solutions and fitted a Gaussian to it. The final value reported for $\nu_{\mathrm o}$ is the median of the fitted Gaussian and the error is its standard deviation. We note that the addition of the GMRT-608 flux densities significantly decreased the uncertainty on the position of the peak frequency for the relevant sources.

\begin{figure*}
  \subfloat{
  \begin{minipage}{80mm}
    \centering
    \includegraphics[width=\columnwidth]{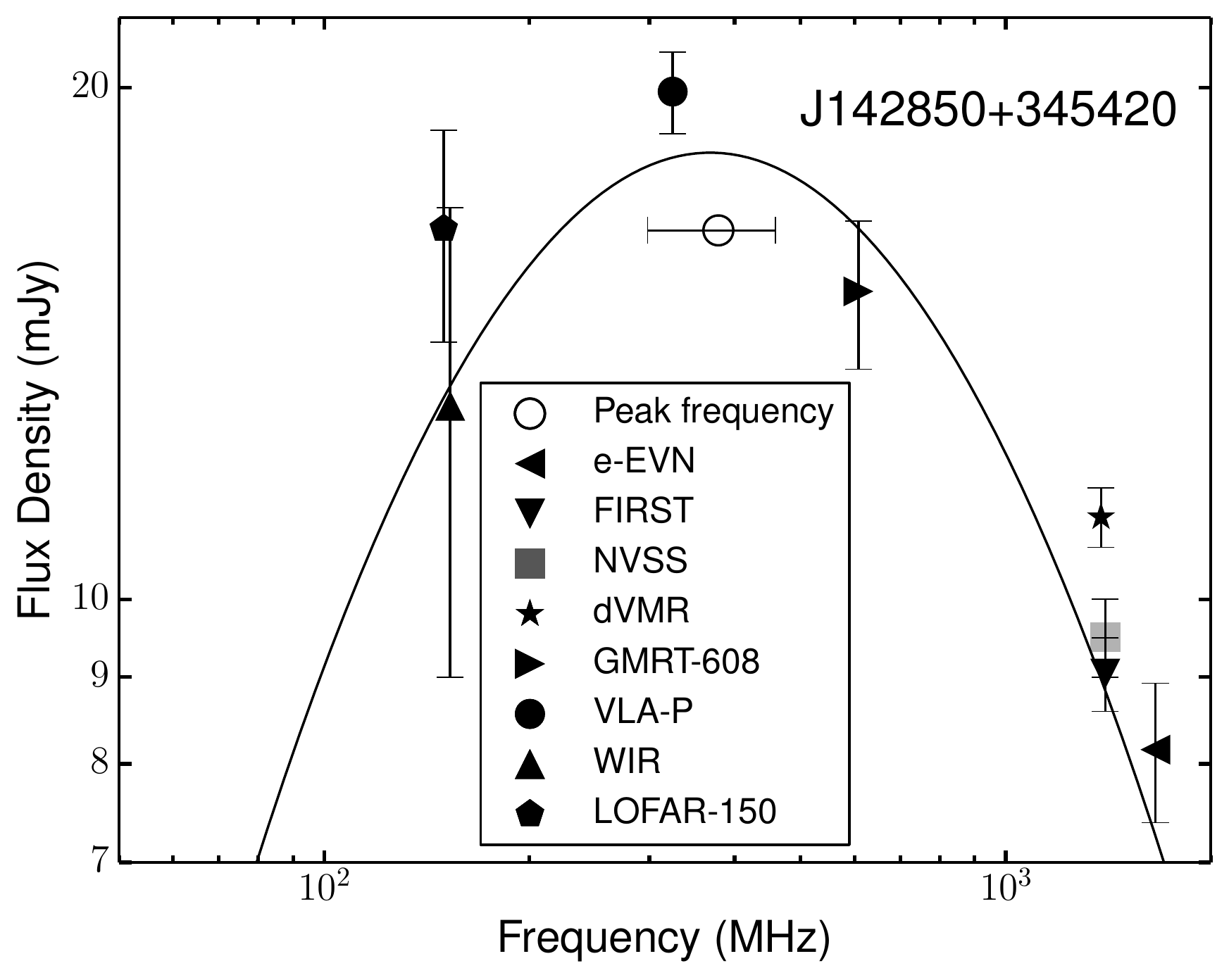}
  \end{minipage}  
  \begin{minipage}{80mm}
    \centering
    \includegraphics[width=\columnwidth]{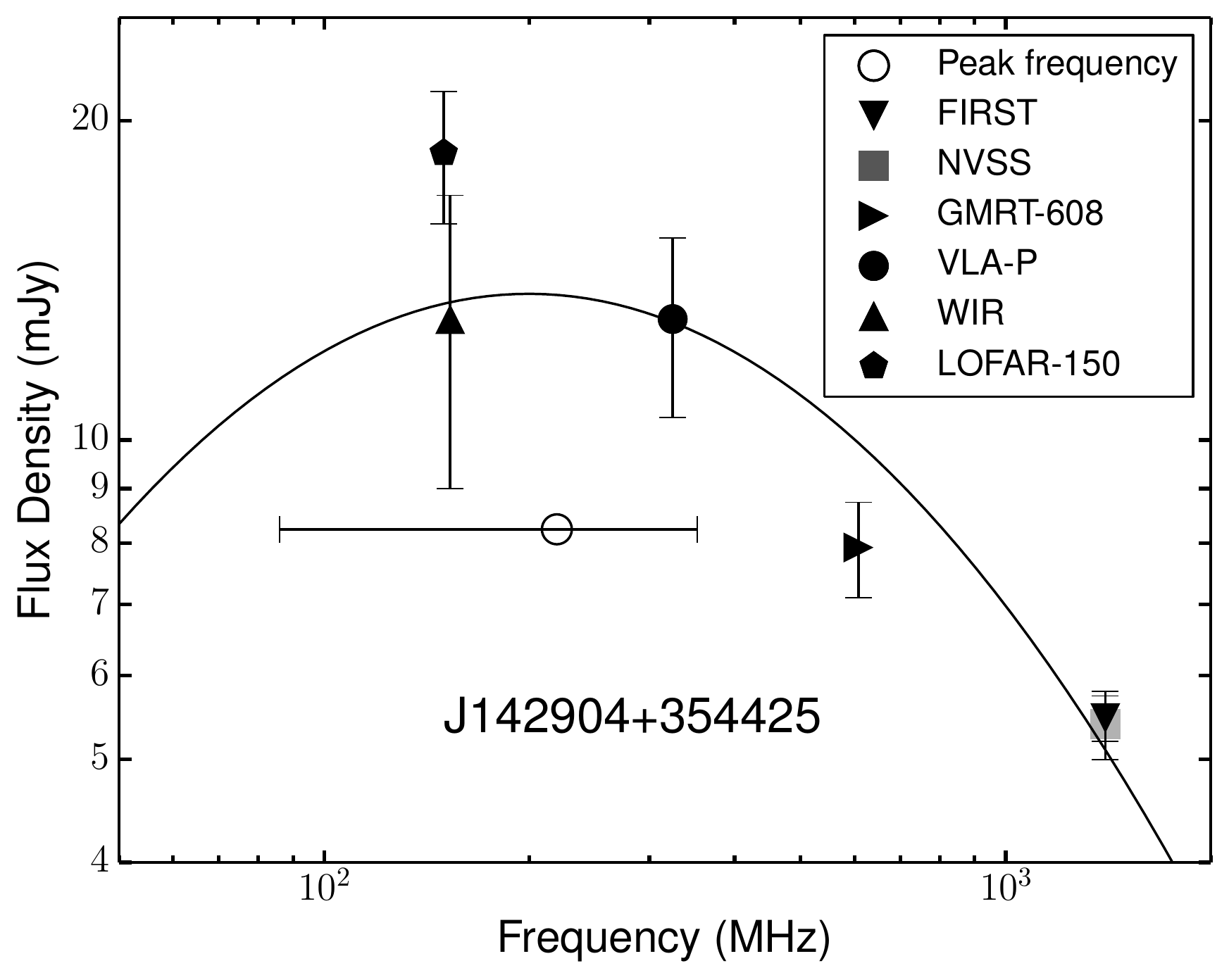}
  \end{minipage}}\\[-2ex] 
  \vspace{0.1cm}
  \subfloat{
  \begin{minipage}{80mm}
    \centering
    \includegraphics[width=\columnwidth]{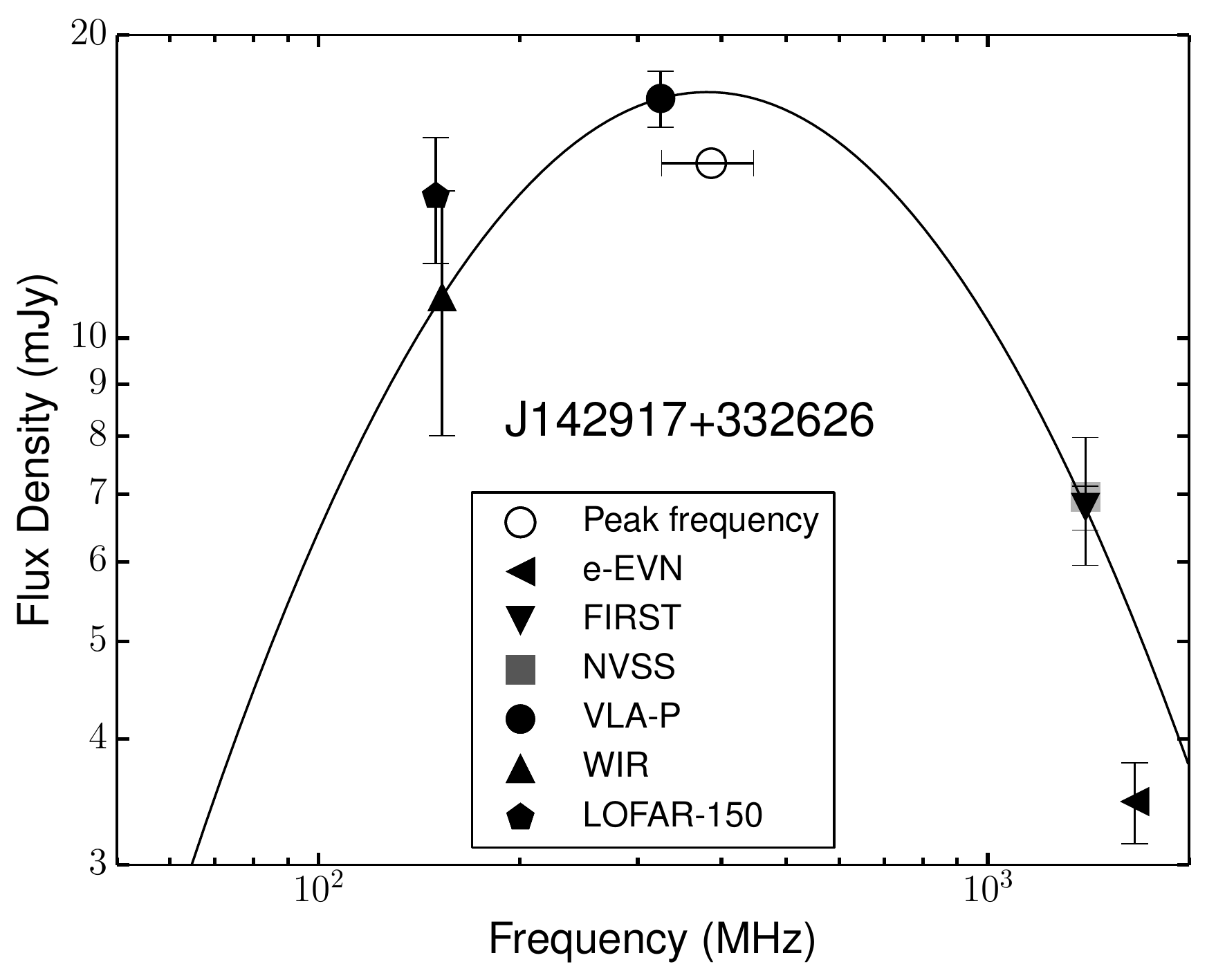}
  \end{minipage}  
  \begin{minipage}{80mm}
    \centering
    \includegraphics[width=\columnwidth]{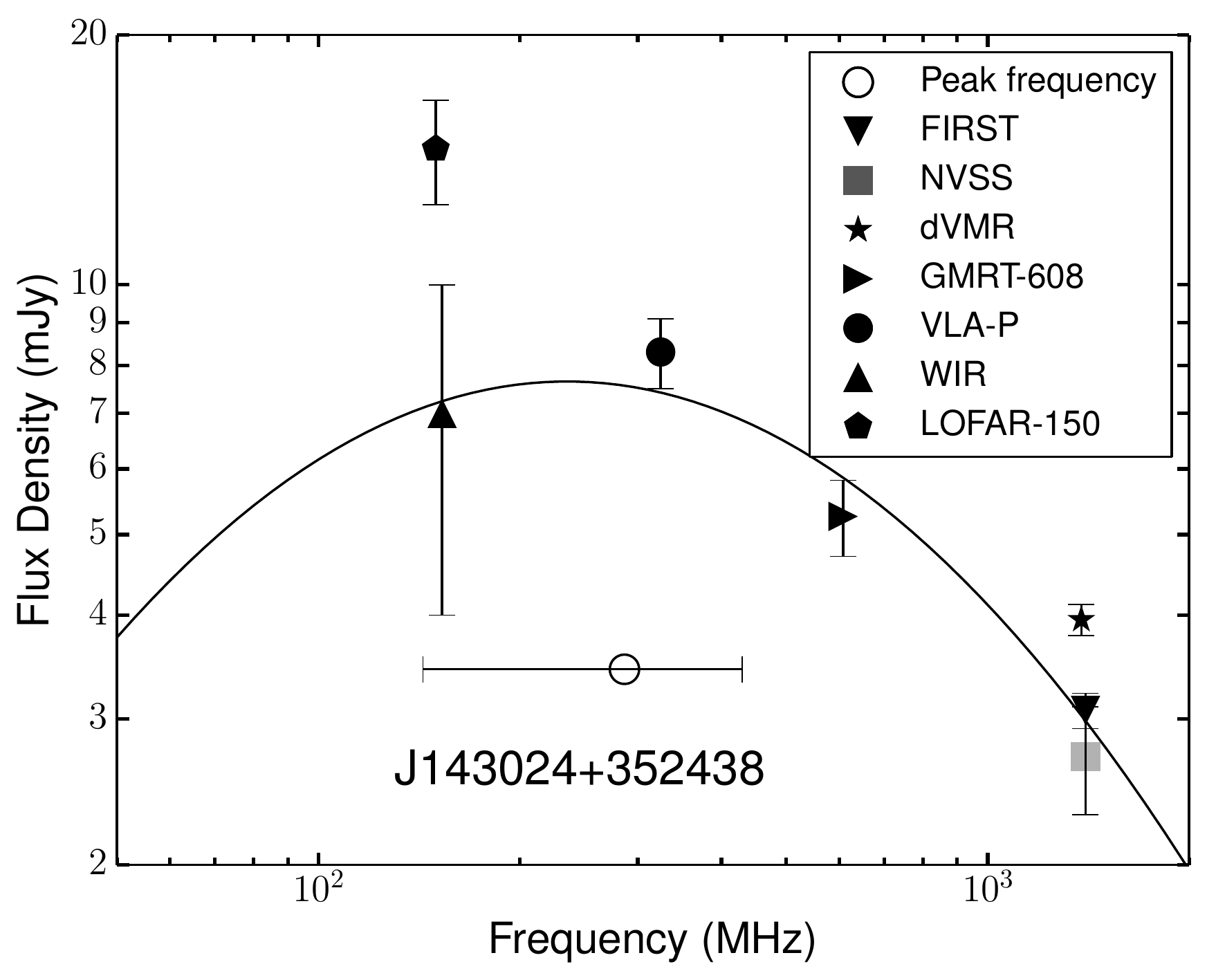}
  \end{minipage}}\\[-2ex]
  \vspace{0.1cm}
  \subfloat{
  \begin{minipage}{80mm}
    \centering
    \includegraphics[width=\columnwidth]{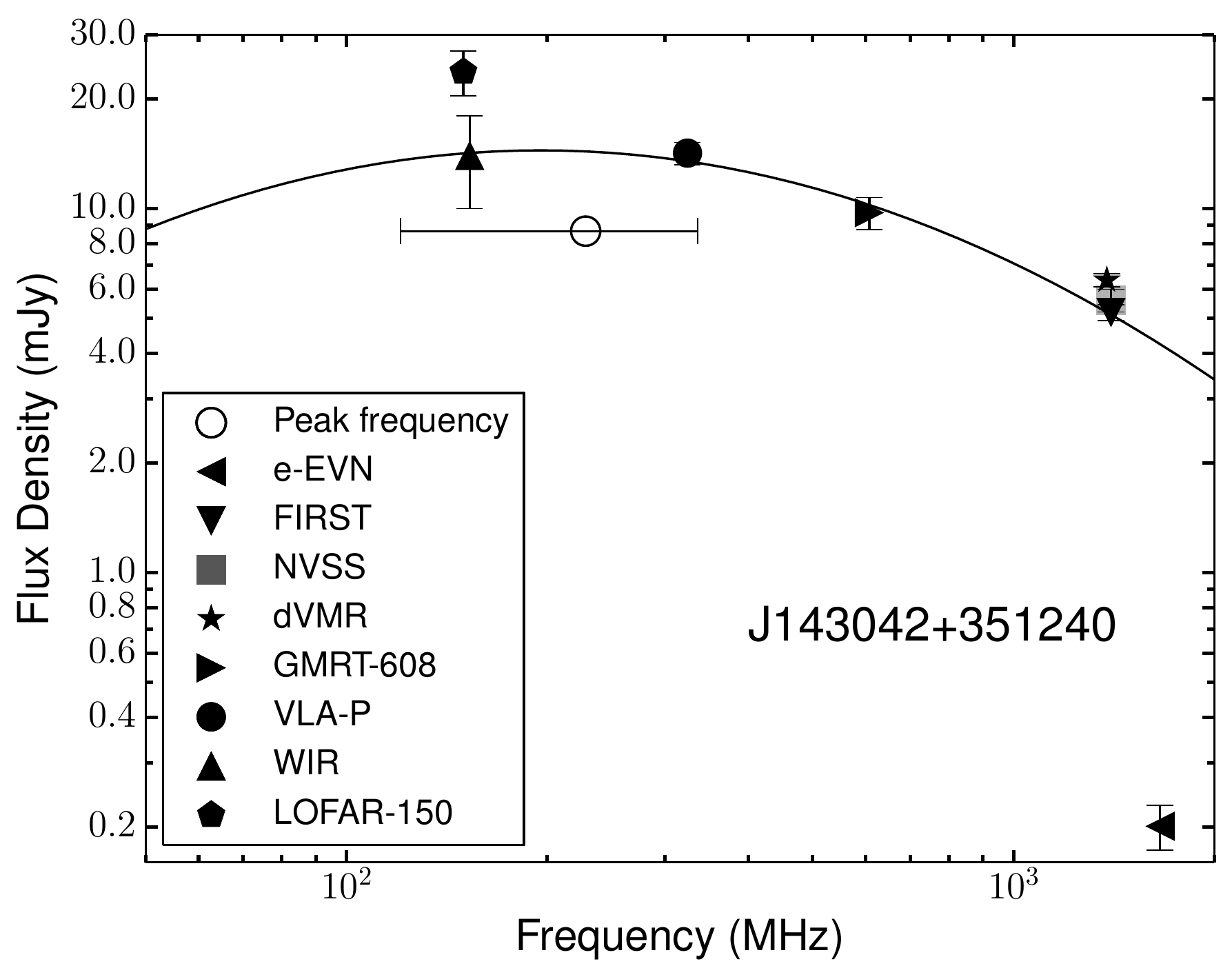}
  \end{minipage}  
  \begin{minipage}{80mm}
    \centering
    \includegraphics[width=\columnwidth]{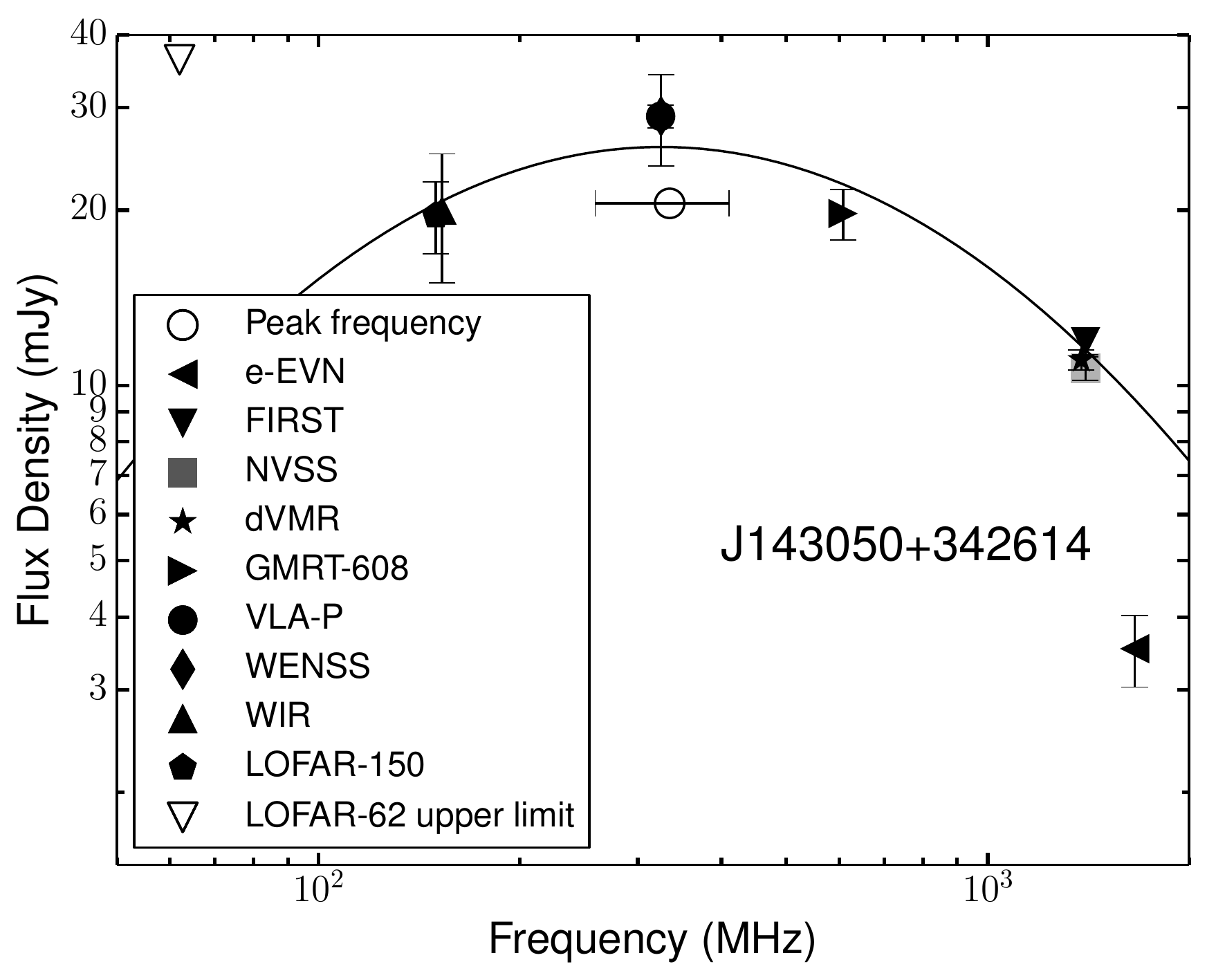}
  \end{minipage}}\\[-2ex]
  \caption{The spectral plots of the sources where the solid line is added to guide the eye. For J143050+342614, the VLA-P and WENSS flux densities are indistinguishable, with the larger of the two error bars being associated with the WENSS flux density. The LOFAR-62 detection limits, given in Table \ref{tbl:lofar-62_detect}, are shown as a empty downward triangle for the sources where they help to constrain the spectrum.}
  \label{fig:spectra}
\end{figure*}

\begin{figure*}
  \subfloat{
  \begin{minipage}{80mm}
    \centering
    \includegraphics[width=\columnwidth]{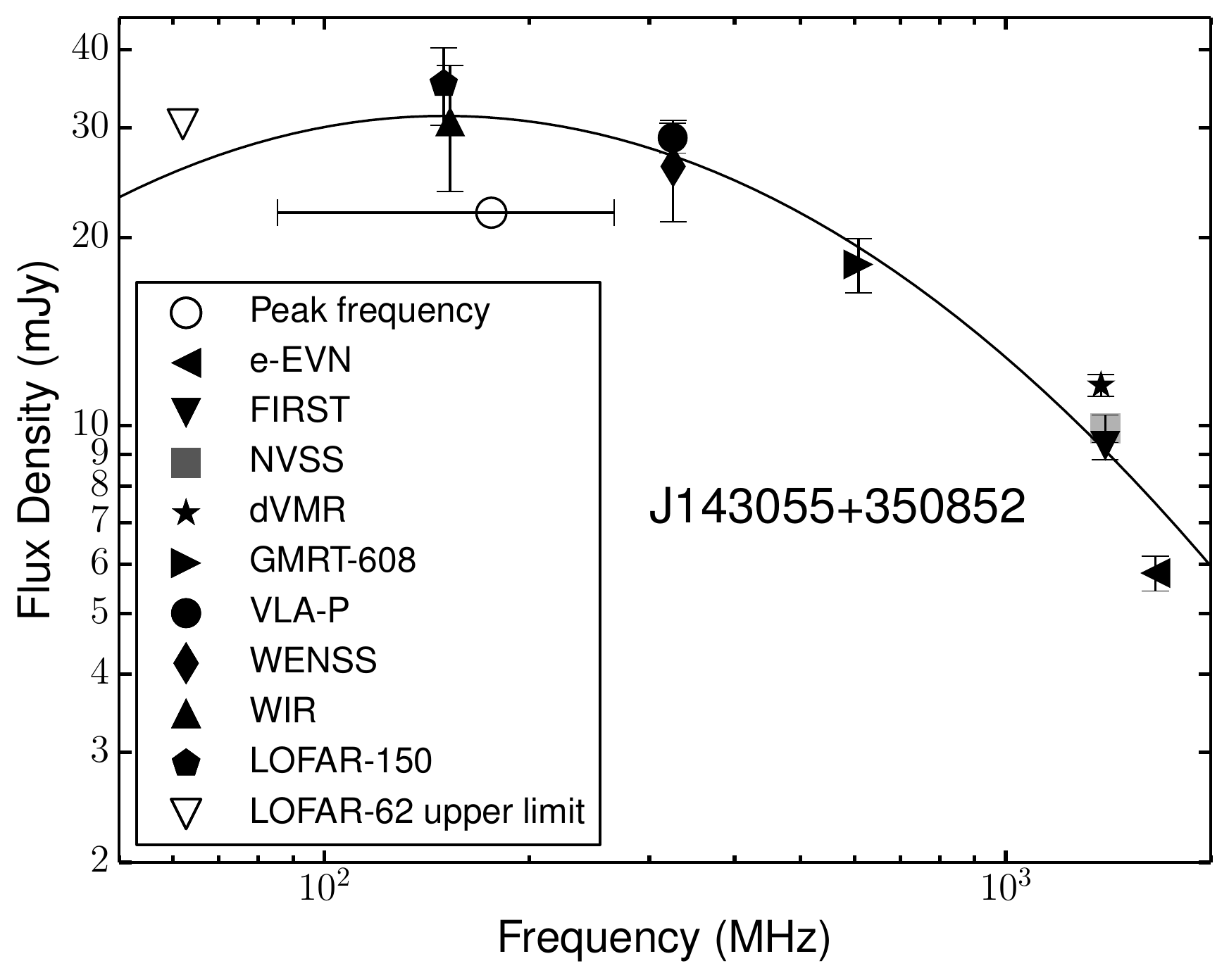}
  \end{minipage}  
  \begin{minipage}{80mm}
    \centering
    \includegraphics[width=\columnwidth]{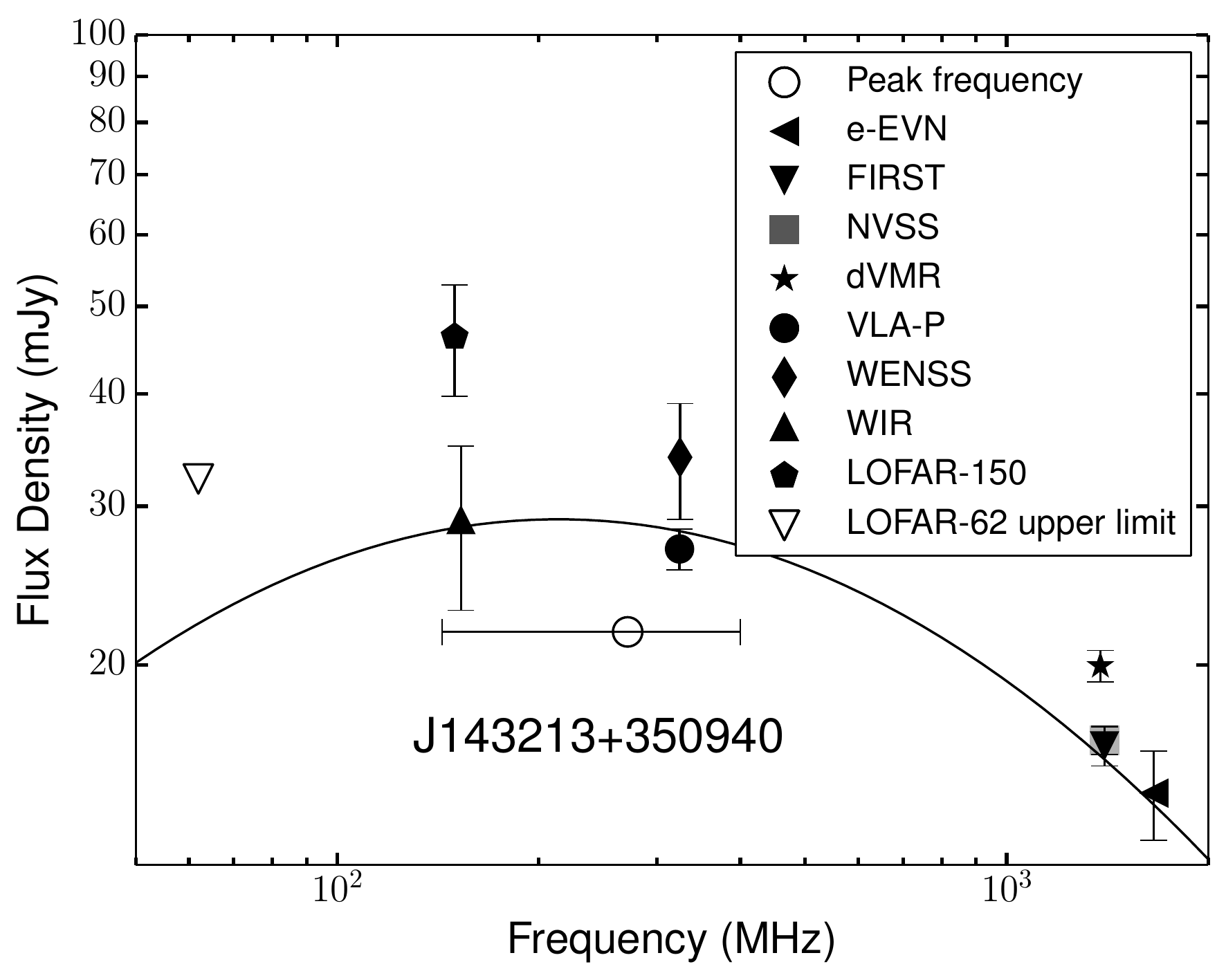}
  \end{minipage}}\\[-2ex]
  \vspace{0.1cm}
  \subfloat{
  \begin{minipage}{80mm}
    \centering
    \includegraphics[width=\columnwidth]{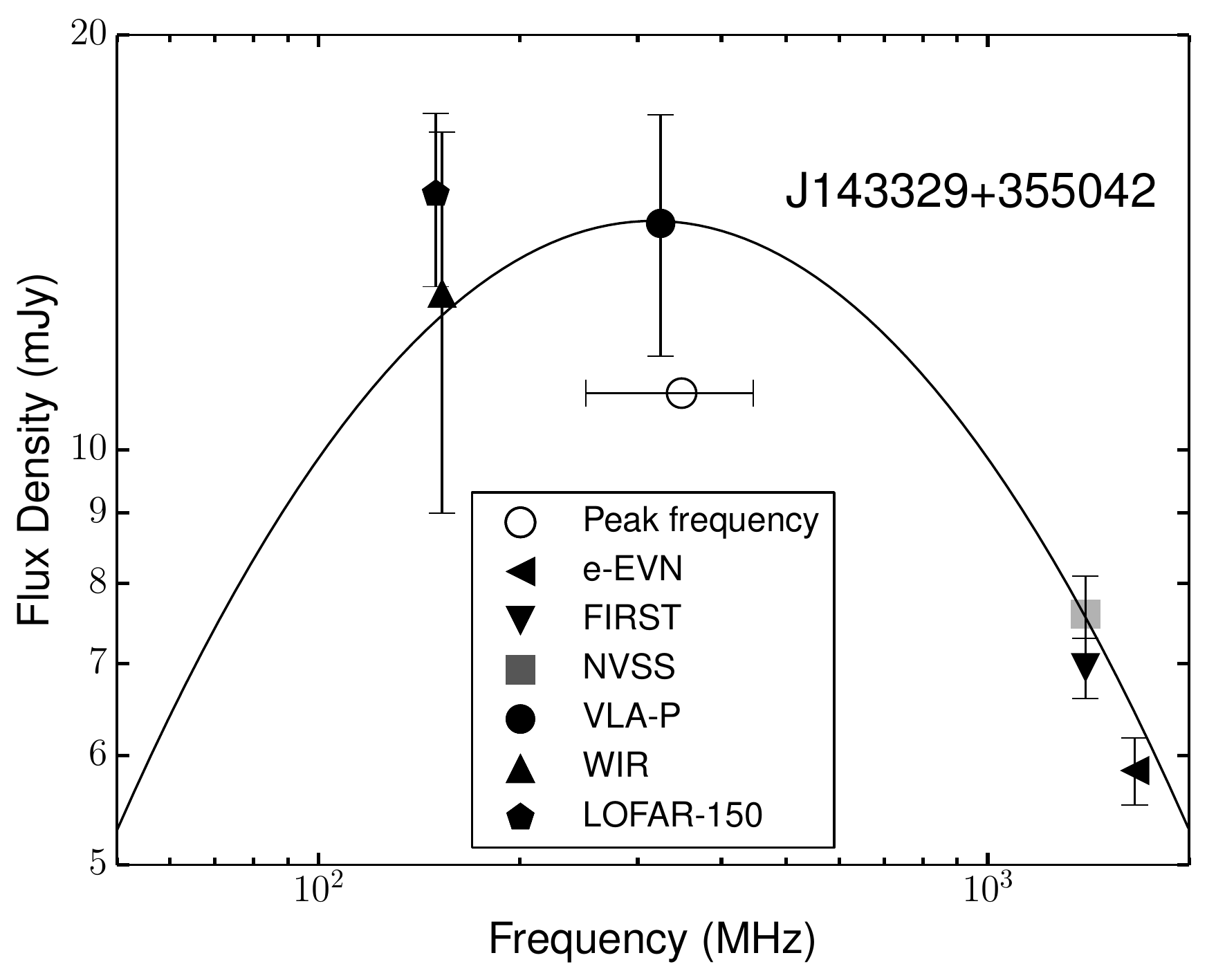}
  \end{minipage}  
  \begin{minipage}{80mm}
    \centering
    \includegraphics[width=\columnwidth]{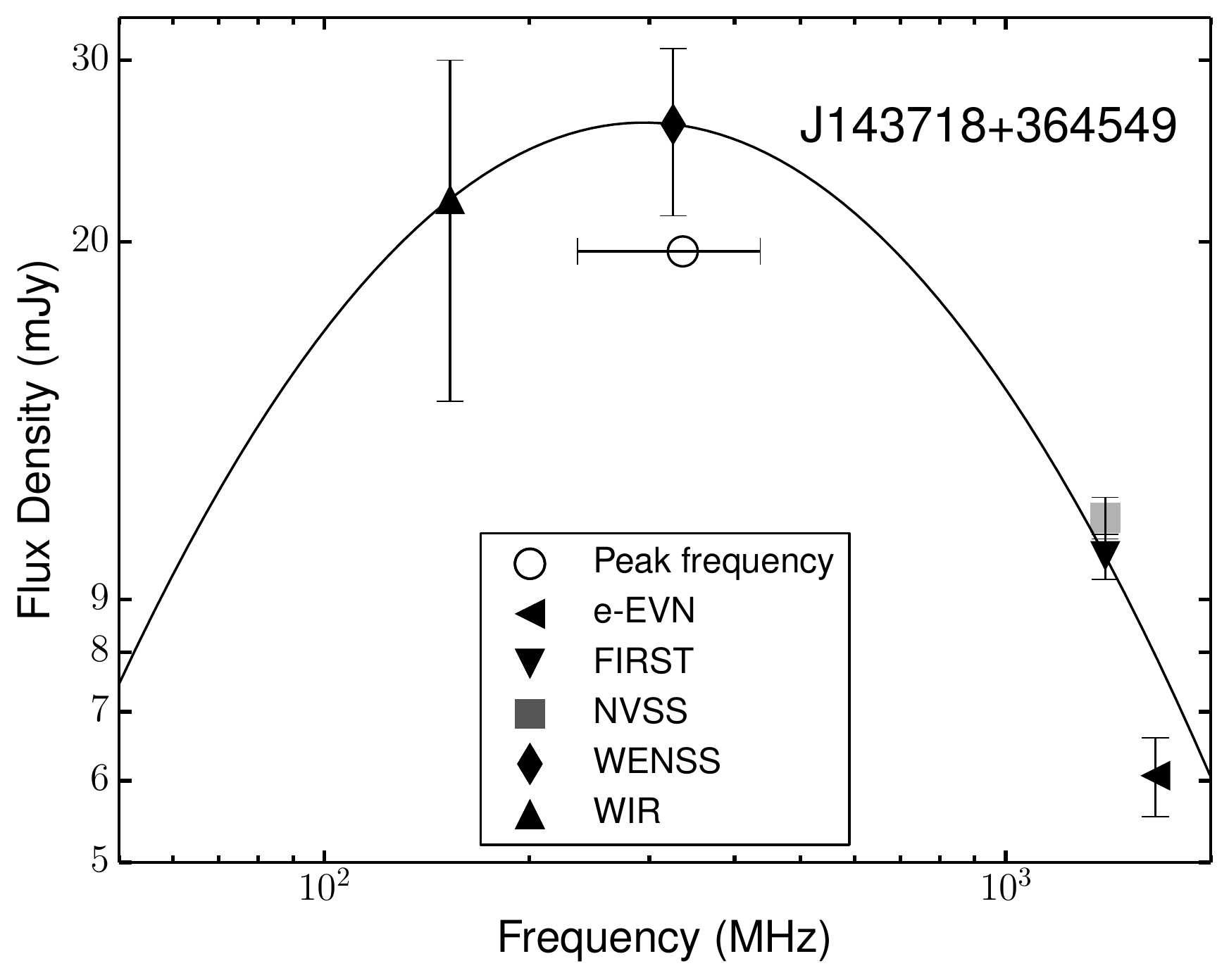}
  \end{minipage}}\\[-2ex] 
  \vspace{0.1cm}
  \subfloat{
  \begin{minipage}{80mm}
    \centering
    \includegraphics[width=\columnwidth]{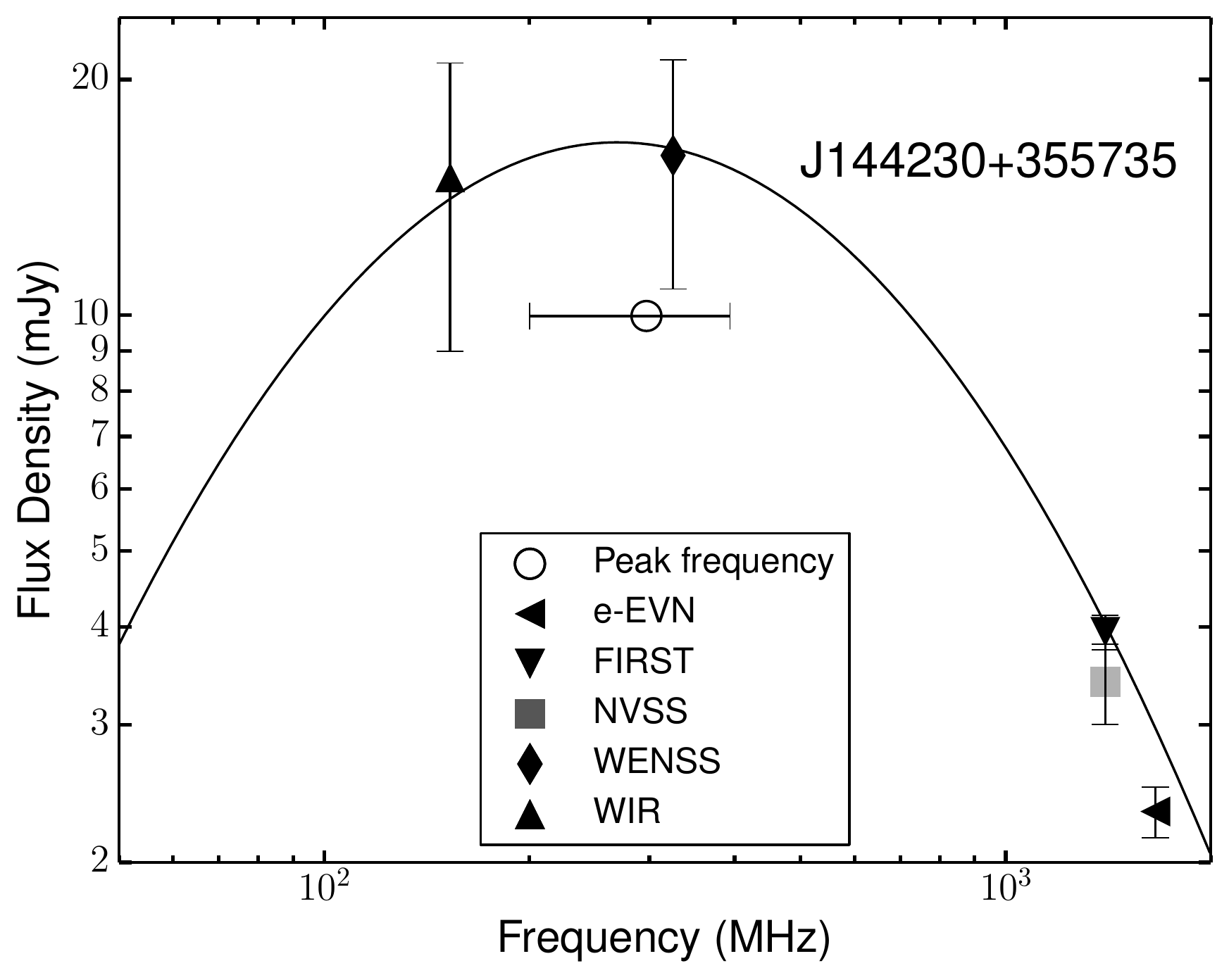}
  \end{minipage}  
  \begin{minipage}{80mm}
  \end{minipage}}\\[-2ex]
  \contcaption{}
\end{figure*}

Column (11) gives the rest-frame turnover frequency of the sources with known redshifts. The values were calculated using $\nu_{\mathrm r} = \nu_{\mathrm o}(1+z)$ where the errors of the $z_{ \mathrm{\textsc{eazy}}}$ values were dealt with as was described for columns (7) and (8). 

Column (12) contains the largest linear sizes (LLS) of the sources calculated using their angular sizes and redshifts. For the sources which have EVN sizes but not redshifts, we calculated upper limits using $z=1$. This can be done because, given a source with fixed linear size, its angular size will decrease as a function of redshift from $z=0$ to $z\sim1$. However, at $z>1$, increasing the redshift of the source results in a slight, but systematic, increase in its angular size (see for example fig. 10 in \citet{falcke2004}). Because of this, for a fixed angular size, a plot of the linear size to which it corresponds as a function of redshift peaks at $z=1$.

\subsection{Are they resolved?}
\label{subsec: resolved}
To test if the source components are resolved in the EVN images, we calculated the minimum resolvable size of each of the Gaussian components fitted to the sources using Equation 2 in \citet{2005AJ....130.2473K}. Since all the fitted source sizes are larger than the minimum resolvable size, we conclude that all the sources and their components are resolved in the EVN images. 

\subsection{1.4\,GHz Variability}
\label{subsec:variability}
In Section \ref{subsec:Source Selection}, we pointed out that J142850+345420, J143024+352438, J143042+351240, J143055+350852 and J143213+350940 have flux density differences of more than 20\,per\,cent between either FIRST and dVMR or NVSS and dVMR. This could potentially indicate that these sources are variable. However, it is surprising that all five sources have flux density differences of more than 20\,per\,cent with the dVMR catalogue, while they all have a flux density difference of less than 12\,per\,cent between FIRST and NVSS. We therefore checked the flux density offset between the dVMR catalogue and FIRST and dVMR and NVSS for the 190 VLA-P sources that we matched to all three catalogues in \citet{coppejans2015}, excluding only the sources that were flagged as extended. Doing this, we found that the median values of $S_{\mathrm{dVMR}}/S_{\mathrm{FIRST}}$, $S_{\mathrm{dVMR}}/S_{\mathrm{NVSS}}$ and $S_{\mathrm{NVSS}}/S_{\mathrm{FIRST}}$ are 1.16, 1.07 and 1.07, respectively. Since the dVMR flux densities were calibrated against NVSS \citep{deVries2002bootes}, it is not surprising that their flux densities agree well with each other. What is interesting is that the flux densities of FIRST and NVSS, the highest and lowest resolution catalogues, agree well with each other but that the values for dVMR and FIRST do not agree as well. If anything we would expect that the dVMR values should be higher than those of FIRST since its resolution is three and a half times lower than that of FIRST. Of the 190 sources, the dVMR flux density of 124 are higher than both those of FIRST and NVSS. This seems to indicate that the dVMR flux densities are slightly higher than those of FIRST for our sub-selection of sources. Hence, the five sources that have a flux density difference of more than 20\,per\,cent between FIRST or NVSS and dVMR are not necessarily variable.

One possible reason why the sources could be variable is that they are blazars, which are radio-loud AGN whose jet is pointed within a small angle of our line of sight \citep[e.g.][]{1999ASPC..159....3U,2013FrPhy...8..609K}. It is however very unlikely that any of the sources (including J142904+354425 and J143024+352438 which we did not detect) are blazars since blazars have brightness temperatures above $\sim10^{10}$\,K \citep{1994ApJ...426...51R,2006ApJ...642L.115H}, which is significantly higher than the values derived for our targets. In addition, blazars have flat spectra in the MHz regime \citep{2013ApJS..207....4M} which is not the case for any of our sources. Hence we do not believe that any of the sources are blazars and are therefore highly unlikely to be variable. It is also worth pointing out that the GPS and CSS sources are the least variable class of radio AGN \citep{o'dea1998}.

\subsection{The origin of the radio emission}
\label{subsec:agn star}
To test if the observed radio emission could be the result of star formation in the host galaxies, we used the same method as \citet{Magliocchetti2014} to differentiate between star forming galaxies and AGN using only radio luminosity. The method presented by \citet{Magliocchetti2014} is based on the results of \citet{McAlpine2013} who used the optical and near-infrared spectral energy distributions of 942 1.4\,GHz radio sources to calculate luminosity functions and redshifts for star forming and AGN dominated radio galaxies. Using these results, \citet{Magliocchetti2014} calculated that the radio power beyond which AGN-powered sources are dominant over star forming sources scale with redshift as 
\begin{equation}
 \log_{10}P_{\mathrm{cross}}(z) = \log_{10}P_{\mathrm{0,cross}} + z
 \label{eq: star agn cut as function of z}
\end{equation}
up to $z\sim1.8$. Where $P_{\mathrm{0,cross}}=10^{21.7}$\,$\mathrm{W\,Hz^{-1}\,sr^{-1}}$ is the value at $z=0$ \footnote{We note that in \citet{Magliocchetti2014} the units for $P_{\mathrm{0,cross}}$ are incorrectly shown as $\mathrm{[W\,Hz\,sr^{-1}]}$.}. \citet{Magliocchetti2014} notes that for a given redshift, the amount of star forming galaxies with radio powers larger than this value drops steeply and that at all redshifts, the radio luminosity function of star-forming galaxies drops off much steeper than that of AGN. Hence, the authors expect there to be very little contamination between the selections of star-forming and AGN galaxies \citep{Magliocchetti2014}.

To calculate the radio power of our sources, we used the same relation as \citet{Magliocchetti2014}:
\begin{equation}
 P_{\mathrm{1.4GHz}} = S_{\mathrm{1.4GHz}}D^2(1+z)^{3-\alpha}.
 \label{eq: radio power}
\end{equation}
Here $P_{\mathrm{1.4GHz}}$ is in units of $\mathrm{[W\,Hz^{-1}\,sr^{-1}]}$, $S_{\mathrm{1.4GHz}}$ is the FIRST flux density in Jy converted to units of $\mathrm{[W\,m^{-2}\,Hz^{-1}\,sr^{-1}]}$, $D$ is the angular diameter distance in meters and $\alpha$ is the high-frequency spectral index of the source reported in Table \ref{tbl:targets}. Note that we use an exponent of $3-\alpha$ as opposed to $3+\alpha$ in Equation \ref{eq: radio power} since \citet{Magliocchetti2014} define the spectral index as $S \propto \nu ^ {-\alpha}$.

Following \citet{Magliocchetti2014}, sources at $z\leq1.8$ are classified as AGN powered if $P_{\mathrm{1.4GHz}}>P_{\mathrm{cross}}(z)$, and as star formation powered if $P_{\mathrm{1.4GHz}}<P_{\mathrm{cross}}(z)$. For sources at $z>1.8$, the classification is done in the same way except that $P_{\mathrm{cross}}(z)$ is always equal to $10^{23.5}\mathrm{W\,Hz^{-1}\,sr^{-1}}$. Using both the minimum \textsc{eazy} redshift allowed inside the errors and the \textsc{lrt} redshift, we found that in all the sources for which we have redshifts, the radio emission is from an AGN not star formation. While we do not have redshifts for J142850+345420, J143718+364549 and J144230+355735, the minimum redshift at which the emission would be the result of star formation is 0.06 for both J142850+345420 and J143718+364549 and 0.1 for J144230+355735.

Brightness temperatures can also be used to differentiate between star formation and AGN activity. Thermal radio emission caused by star formation typically has $T_{\mathrm{b}}<10^5$\,K \citep{Sramek1986,Condon1991,Kewley2000} while $T_{\mathrm{b}} \geq 10^6$\,K can be used as an indicator of non-thermal emission from AGN \citep[e.g.][]{Kewley2000,Middelberg2011}. Since all the sources have $T_b \geq 10^6$\,K inside their errors, this confirms that the radio emission is from AGN activity. This is further supported by the morphologies of the sources that are resolved into multiple components (Section \ref{sec:Comments on Individual Sources}). We note that non-thermal emission could also originate from a supernova remnant or a nuclear supernova remnant complex \citep[e.g.][]{2012MNRAS.423.1325A}. However, this possibility is excluded by the power cut described above.


\section{Comments on Individual Sources}
\label{sec:Comments on Individual Sources}

\subsection{Non-detections}  
The following two sources were not detected in the EVN images. Tapering the data also did not result in a detection, and from Section \ref{subsec:agn star} we do not expect that the radio emission is caused by star formation. Below we conclude that the non-detections are because the sources have extended lobes which were resolved out by the high-resolution EVN observations while the faint core remains undetected. We note that this conclusion could be tested by observing the sources with an instrument that probes angular sizes between 0.15 and 5\,arcsec.

\subsubsection{J142904+354425}
\label{subsec:J142904+354425}
J142904+354425 is not detected in the naturally weighted, $5\times5$\,arcsec EVN image with $23\mu \mathrm{Jy\,beam^{-1}}$ noise and a $30\times34$\,mas restoring beam. J142904+354425 was included in our MPS selection in \citet{coppejans2015} and was not flagged as being variable based on its FIRST and NVSS flux densities. No match could be found for it in the dVMR catalogue because it lies outside the area that was imaged. J142904+354425 was detected by \citet{Garrett2005} using the WSRT at 1.4\,GHz and found to be smaller than 5\,arcsec, having an integrated flux density of $3.026\pm0.026$\,mJy\footnote{We note that \citet{Garrett2005} did not include the absolute calibration error (which the authors estimate to be less than 2\,per\,cent) in their flux density error.} which is only 45\,per\,cent of the FIRST value. Hence we cannot rule out the possibility that J142904+354425 is variable, but based on the arguments at the end of Section \ref{subsec:variability}, we consider this to be very unlikely. 

To constrain J142904+354425 size we note that it has a deconvolved major and minor axis FWHM of $2.8\pm0.5$ and $1.7\pm0.5$\,arcsec, respectively, in FIRST at a position angle of $41.6^{\circ}$ east of north. We can estimate its minimum size by assuming it consists of a single component that has a surface brightness below the EVN's detection threshold. Taking the detection threshold to be five times the rms noise of the image ($0.115\mathrm{\,mJy\,beam^{-1}}$), the flux density has to be spread out over at least $3.026/0.115 = 26.3$ beams which translates to a minimum angular size of approximately 0.9\,arcsec.

Considering that J142904+354425's predicted flux density at 1.7\,GHz is $4.9\pm0.4$\,mJy, that it is an AGN (Section \ref{subsec:agn star}) and non-variable, we conclude that the non-detection is because J142904+354425 was resolved out by the EVN observations. Finally, as is evident from the Fig. \ref{fig:spectra}, we cannot say for certain that J142904+354425's spectrum turns over.

\subsubsection{J143024+352438}
\label{subsec:J143024+352438}
J143024+352438 is not detected in the naturally weighted, $5\times5$\,arcsec EVN image with $11\mu \mathrm{Jy\,beam^{-1}}$ noise and a restoring beam of $6\times18$\,mas. Considering that J143024+352438 has a predicted 1.7\,GHz flux density of $2.7\pm0.2$\,mJy, we should have easily detected it if it is a compact source. From Fig. \ref{fig:spectra} it is clear that the flux density difference between the WIR and LOFAR-150 catalogues ($7.0\pm3.0$ and $14.6\pm1.5$\,mJy, respectively) can have a significant impact on the shape of J143024+352438's spectrum. It is worth pointing out that J143024+352438 has a signal-to-noise ratio (SNR; defined as the peak brightness divided by the local rms noise) of 49 and 5 in the LOFAR-150 and WIR images, respectively. It is therefore possible that the spectrum can be described by a single power law between 150 and 1400\,MHz.  

Since we do not expect that the radio emission is caused by star formation\footnote{We note that matching J143024+352438 to the SDSS Data Release 10 catalogue \citep[https://www.sdss3.org/dr10/;][]{2014ApJS..211...17A}, we found a source (SDSS J143025.19+352441.3) 3.3\,arcsec away from J143024+352438 with a photometric redshift of $0.342\pm0.1167$. This could influence whether J143024+352438 is classified as being dominated by star formation or an AGN. However, using the analysis in Section \ref{subsec:agn star}, J143024+352438 would only be classified as being dominated by star formation if it is at a redshift below 0.11. }, we conclude that J143024+352438 has lobes which were resolved out by the EVN observations while the faint core remains undetected. J143024+352438's maximum size is set by its deconvolved major axis FWHM of $4.2\pm0.8$\,arcsec in FIRST, which is at a position angle of $164.7^{\circ}$. J142904+354425's minor axis is unresolved in FIRST. Using the same argument presented for J142904+354425, we calculated a minimum angular size of approximately 1\,arcsec for J143024+352438.

\subsection{Marginally resolved sources}
The following four sources appear as single components in the EVN images that do not have a discernible structure. Note that from Section \ref{subsec: resolved}, all of the sources are resolved. Below we discuss each of the sources individually, focusing on, among other things, the percentage of the predicted flux density that was recovered from the image and the source variability. If the percentage of the predicted flux density recovered from the image is low, it could indicate that the measured source size is not a good estimate of the true source size. If, however, the sources is variable, it would mean that the predicted flux density is unreliable. Hence the source could be more extended or has an additional component that could have been missed in the EVN image. While we argue that none of the sources are variable, based on the percentage of the predicted flux densities recovered from the images, we conclude that J143042+351240 and J143050+342614 likely have undetected structure, and are therefore larger than indicated in Table \ref{tbl:derived_info}.

\subsubsection{J142917+332626}
\label{subsec:J142917+332626}
J142917+332626 is present in both the NVSS and dVMR image cutouts, but not in the catalogues because there is another source $\sim35$\,arcsec away which blends with it. To determine its flux density in NVSS we simultaneously fit it and the nearby source using the \textsc{pybdsm} source detection package\footnote{http://dl.dropboxusercontent.com/u/1948170/html/index.html}. From this we found that it has an integrated flux density of $6.96\pm1.01$\,mJy, which differs by 2.5\,per\,cent from the FIRST value. J142917+332626 was also observed by \citet{Ciliegi1999} at 1.4\,GHz with the VLA in C configuration. The authors found a flux density of $6.27\pm0.04$\,mJy for J142917+332626 which differs by 11\,per\,cent from the FIRST value. Considering that we recovered between 51 and 63\,per\,cent of the predicted flux density, we expect that the reported size for J142917+332626 is a good estimate of the true size.

Matching the VLA-P sources to the Chandra XBo\"{o}tes X-ray survey of the Bo\"{o}tes field \citep{Murray2005} with a search radius of 10\,arcsec, J142917+332626 was the only source from our 11 sources for which we found a counterpart. The centroid of the matched source, CXOXB J142917+332626.4, is 0.32\,arcsec away from the VLA-P source and was detected in all three of Chandra's bands with seven, one and six counts in the 0.5--7, 0.5--2 and 2--7\,keV bands, respectively. We note that, since J142917+332626 can unambiguously be classified as an MPS source (Fig. \ref{fig:colour-colour}), it is one of the few turnover sources with a X-ray counterpart.

Matching the FIRST sources to those in XBo\"{o}tes, \citet{Bouchefry2009} associated CXOXB J142917+332626.4 with the same FIRST source to which we matched the VLA-P source. In addition, CXOXB J142917+332626.4 was matched to an optical source in the NOAO Deep Wide-Field Survey (NDWFS) survey of the field by \citet{Brand2006}. Using the publicly available \textsc{hyperz}\footnote{http://webast.ast.obs-mip.fr/hyperz/} code along with the optical information, \citet{Bouchefry2009} determined a photometric redshift of $0.960^{+1.227}_{-0.806}$ for J142917+332626. This value is consistent with the \textsc{eazy} redshift of J142917+332626 and is lower than the \textsc{lrt} value.

Based on its high X-ray to optical flux ratio, $\log_{10}(f_{(2-7)\mathrm{keV}}/f_{\mathrm{opt}}) = 1.37$, J142917+332626 is expected to either be a high-redshift source and/or dust obscured \citep[][and references therein]{Bouchefry2009}. The author classifies it as an obscured AGN (AGN-2) using its hardness ratio and X-ray luminosity.

\subsubsection{J143042+351240}
\label{subsec:J143042+351240}
J143042+351240's EVN flux density of 0.2\,mJy is only $4\pm1$\,per\,cent of the predicted value. Tapering the data did not increase J143024+352438's flux density significantly. Since the emission is related to AGN activity and J143042+351240 is not variable (Sections \ref{subsec:agn star} and \ref{subsec:variability}), this leads us to conclude that the true size of J143042+351240 is larger than indicated in columns (7) and (12) in Table \ref{tbl:derived_info}. If this is the case, the remaining flux density could be in low surface brightness emission surrounding the source. In this case the true source size will be larger than the measured value, but likely not significantly. If however the missing flux density is in a second component, the size could be a significant underestimate of the true source size.

We were able to match J143042+351240 to a source in the AGN and Galaxy Evolution Survey \citep[AGES;][]{Kochanek2012} catalogue which measured redshifts for 23,745 galaxies and AGN in the Bo\"{o}tes field. The photometric redshift of the AGES source (0.96) agrees very well with the values reported in Table \ref{tbl:targets}. The flux density difference between the WIR and LOFAR-150 catalogues ($14.0\pm4.0$ and $23.8\pm2.4$\,mJy, respectively) can have a significant impact on the shape of J143042+351240's spectrum (Fig. \ref{fig:spectra}). The two catalogues have SNRs of 9 and 98, respectively. It is therefore possible that the spectrum (Fig. \ref{fig:spectra}) can be described by a single power law between 150 and 1400\,MHz. 

\subsubsection{J143050+342614}
\label{subsec:J143050+342614}
We only recovered 33\,per\,cent of the predicted flux density for J143050+342614. This percentage did not increase when we applied an uv-taper to the image. The emission is related to AGN activity and the source is not variable (Sections \ref{subsec:agn star} and \ref{subsec:variability}). As is the case with J143042+351240, this leads us to conclude that J143050+342614 is larger than indicated in columns (7) and (12) in Table \ref{tbl:derived_info}. We note that J143050+342614 can unambiguously be classified as an MPS source when including the errors on its spectral indices. It is also matched to a WENSS source with the WENSS and VLA-P flux densities differing by less than one per cent. This results in the two points being indistinguishable in J143050+342614's spectral plot (Fig. \ref{fig:spectra}) with the larger of the two error bars being associated with the WENSS flux density.

\subsubsection{J143329+355042}
\label{subsec:J143329+355042}
J143329+355042 is composed of a single component in the EVN image and lies outside the survey area covered by the dVMR catalogue. We recovered $92\pm9$\,per\,cent of J143329+355042's predicted flux density. This indicates that its measured size is a good estimate of its true size. J143329+355042's spectrum (shown in Fig. \ref{fig:spectra}) flattens toward lower frequencies and could turn over (Table \ref{tbl:targets}).

\subsection{Resolved sources}

The following five sources are resolved in the EVN images and have a discernible structure. Below we discuss each of the sources individually and argue that the observed structures are lobes and/or hotspots in their jets.

\subsubsection{J142850+345420}
\label{subsec:J142850+345420}
\begin{figure}
  \includegraphics[width=\columnwidth]{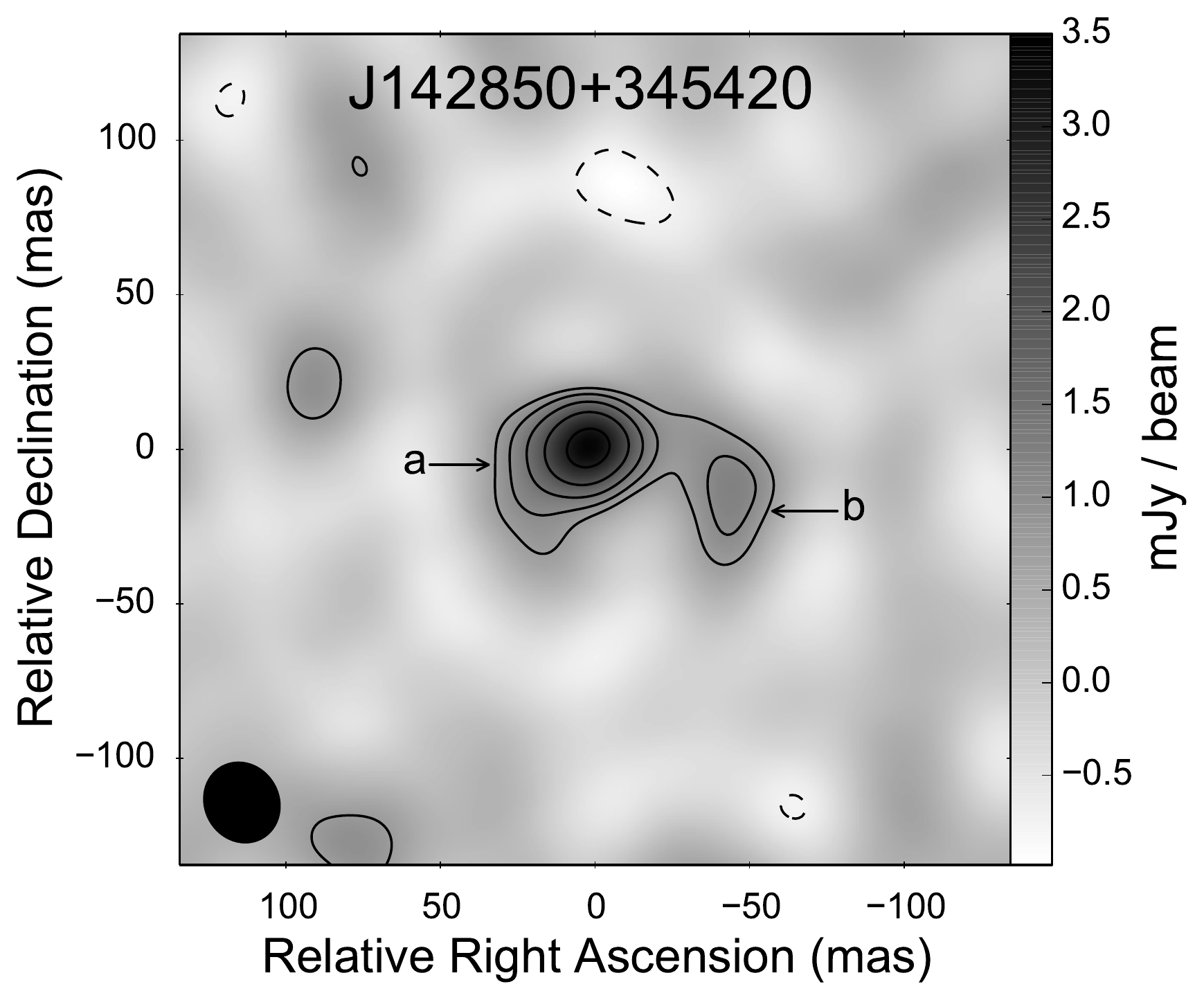}
  \caption{Naturally weighted EVN image of J142850+345420 that was made using an uv-taper. The restoring beam is shown in the bottom left corner and has a size of $24.1\times26.8$\,mas at a major axis position angle of $29\fdg4$. The contours are drawn at $-3$ and 3 times the image noise, increasing in factors of $\sqrt 2$ thereafter.}
  \label{fig:J142850+345420}
\end{figure}

J142850+345420, shown in Fig. \ref{fig:J142850+345420}, has a double structure in the EVN image with J142850+345420a and J142850+345420b being detected at a $14\sigma$ and $5\sigma$ level, respectively. Fig. \ref{fig:J142850+345420} was made using natural weighting and applying an uv-taper. The full-resolution, uniformly weighted image has a typical resolution of $3\times10$\,mas, but we could not confirm the detection of J142850+345420b. Applying the uv-taper increased the beam size to $24.1\times26.8$\,mas and resulted in the flux density of both components increasing. Specifically, the flux density of J142850+345420a increased by a factor of 1.7 and the percentage of recovered flux density of the source as a whole increases from 44 to 99\,per\,cent. This is a clear indication that both components were resolved in the full resolution image.

For J142850+345420, we recovered 99\,per\,cent of the predicted flux density, this indicates that its measured size is a good estimate of its true size. J142850+345420 was however flagged as being variable because of a 26\,per\,cent flux density difference between the FIRST and dVMR catalogues. Considering that the NVSS and FIRST flux densities only differ by 7\,per\,cent, and that, as discussed in Section \ref{subsec:Source Selection}, the dVMR flux densities seem to be systematically higher than the FIRST and NVSS values for our sample of sources, it is unlikely that J142850+345420 is variable.
 
J142850+345420 lies inside the multiwavelength optical coverage of the Bo\"{o}tes field, but no match could be found for the source. Consequently, we could not determine a photometric redshift for it. This likely indicates that J142850+345420 is at $z>2$. If this is the case, J142850+345420 is AGN dominated since we calculated in Section \ref{subsec:agn star} that the radio emission would be the result of AGN activity if the source is at $z>0.06$. Since we do not have a redshift for J142850+345420, we calculated the brightness temperatures for J142850+345420a,b using a redshift of zero. The values are therefore lower limits and allow us to conclude that J142850+345420 is consistent with being an AGN with a core-jet structure. 

Since we do not have a redshift for J142850+345420, we calculated the brightness temperatures for J142850+345420a,b using a redshift of zero. The values are therefore lower limits and allow us to conclude that J142850+345420 is consistent with being an AGN with a core-jet structure. We also note that, J142850+345420 lies inside the multiwavelength optical coverage of the Bo\"{o}tes field, but no match could be found for the source. Consequently, we could not determine a photometric redshift for it. This likely indicates that J142850+345420 is at $z>2$. It also indicates that J142850+345420 is AGN dominated since we calculated in Section \ref{subsec:agn star} that the radio emission would be the result of AGN activity if the source is at $z>0.06$.  

\subsubsection{J143055+350852}
\label{subsec:J143055+350852}
\begin{figure}
  \includegraphics[width=\columnwidth]{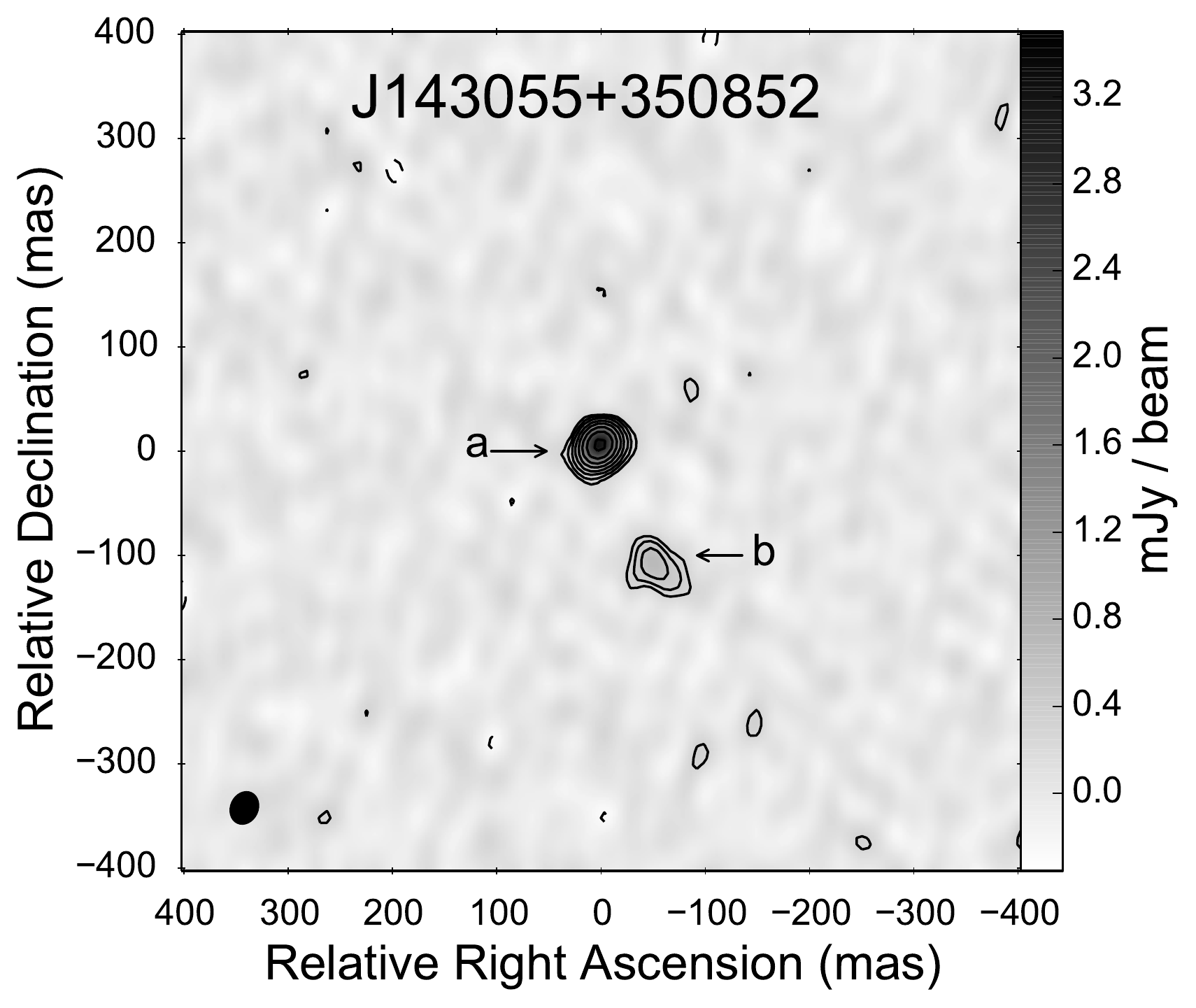}
  \caption{Uniformly weighted EVN image of J143055+350852. The restoring beam is shown in the bottom left corner and has a size of $26.4\times32.1$\,mas at a major axis position angle of $-24\fdg5$. The contours are drawn at $-3$ and 3 times the image noise, increasing in factors of $\sqrt 2$ thereafter.}
  \label{fig:J143055+350852}
\end{figure}

J143055+350852, shown in Fig. \ref{fig:J143055+350852}, is composed of two components. The \textsc{eazy} and \textsc{lrt} redshift values for J143055+350852 differ significantly. Despite this, the brightness temperature of J143055+350852a is consistent with it being an AGN. Considering that J143055+350852b is extended in the direction parallel to the line between the components and that it is fainter than J143055+350852a, we suggest that J143055+350852 has a core--jet structure. Since we recovered most of the predicted flux density ($71\pm6$\,per\,cent), and the remaining flux density is likely located either between the two components or very near them, we do not expect that J143055+350852 is significantly more extended than indicated in Table \ref{tbl:derived_info}. From Fig. \ref{fig:spectra}, it is clear that the upper limit on the LOFAR-62 flux density indicates that the spectrum is turning over.

\subsubsection{J143213+350940}
\label{subsec:J143213+350940}
\begin{figure}
  \includegraphics[width=\columnwidth]{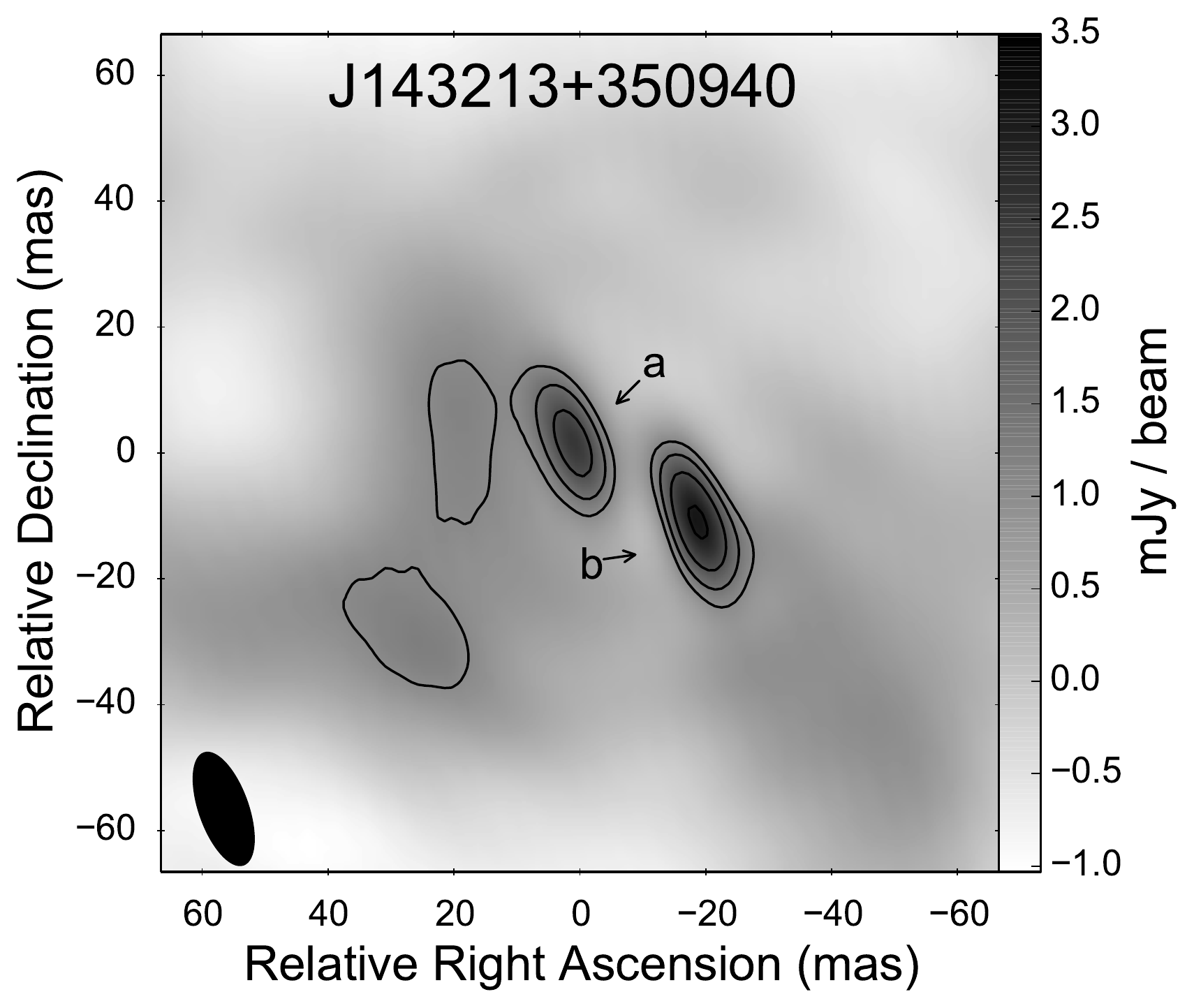}
  \caption{Uniformly weighted EVN image of J143213+350940 without uv-tapering showing a likely CSO structure. The restoring beam is shown in the bottom left corner and has a size of $7.8\times18.9$\,mas at a major axis position angle of $19\fdg4$. The contours are drawn at $-3$ and 3 times the image noise, increasing in factors of $\sqrt 2$ thereafter.}
  \label{fig:J143213+350940}
\end{figure}

The uniformly weighted EVN image of J143213+350940 (Fig. \ref{fig:J143213+350940}) has a likely symmetric double structure. Both components have brightness temperatures consistent with emission from AGN activity. After imaging J143213+350940 with uniform weighting, we also imaged it using natural weighting with an uv-taper. Note that the information given for J143213+350940 in Table \ref{tbl:derived_info} was derived from the naturally weighted, uv-tapered image, while the information for J143213+350940a,b was derived using the uniformly weighted image. Because the beam size of the uv-tapered image is $24\times29$\,mas and J143213+350940a,b are only separated by $\sim31$\,mas, their flux densities merge in the uv-tapered image. Fitting the resulting single source with a single elliptical Gaussian, we found it to have a FWHM of nearly 70\,mas and flux density of 14.41\,mJy. This is $\sim6$\,mJy more than the sum of the flux densities of the individual components. Hence, while the source is resolved, it is clear that there is a significant amount of extended emission between and surrounding the components. From the uv-tapered image, we recover $94\pm12$\,per\,cent of the predicted flux density. Hence, the size measured from the uv-tapered image is a good estimate of J143213+350940's true size.

From Fig. \ref{fig:spectra} it is clear that the VLA-P and WENSS flux densities differ ($26.9\pm1.4$ and $34.0\pm5.0$, respectively), but are nearly consistent within their $1\sigma$ errors. If we take the VLA-P flux density to be the correct value at 325\,MHz, J143213+350940 is not an MPS source (Fig. \ref{fig:colour-colour}) whose spectrum either flattens towards lower frequencies or can be described by a single power law between 153 and 1400\,MHz (Table \ref{tbl:targets}). This means that the turnover frequencies and linear sizes reported in Table \ref{tbl:derived_info} will not be good estimates of the true values. However, if we take the WENSS flux density to be the correct value, $\alpha_{\mathrm{low}}$ and $\alpha_{\mathrm{high}}$ changes to $0.2\pm0.3$ and $-0.5\pm0.1$, respectively. This means that J143213+350940 is an MPS source as is shown by the point labeled ``J143213+350940 WENSS'' in Fig. \ref{fig:colour-colour}. The percentage of the predicted flux density that was recovered from the image changes to $97\pm13$\,per\,cent while $\nu_o=337\pm75$\,MHz and $\nu_{\mathrm r}=666\pm152\,\&\,660\pm146$\,MHz in Table \ref{tbl:derived_info}.

To try to find a reason for the flux density difference between the VLA-P and WENSS catalogues, we double checked both catalogues and images of the source. Since the nearest neighboring source is more than 3\,arcmin away, J143213+350940 is unresolved in the WENSS, VLA-P, FIRST and NVSS images, and there is nothing in the images that can explain the difference. One possibility is that J143213+350940 is intrinsically variable. However, as discussed in Section \ref{subsec:variability}, we consider this to be very unlikely. Another possibility is that the difference could be caused by (as a whole or in part) interplanetary or interstellar scintillation. Typically, interstellar scintillation results in flux density variations of a few per cent \citep[e.g.][]{2000ApJ...539..300S,2005ApJS..159..242G}, which is much less than the 21\,per\,cent difference observed here. In addition, the VLA-P observations were carried out over four days during a period of more than a month \citep{coppejans2015}, while each field in WENSS was observed in 18 snapshots spread over a 12-h period \citep{wenss}. This averaged out the variations, which have maximum typical time scales of about an hour \citep{1998ApJ...507..846C,2005ApJS..159..242G}.

The WIR and LOFAR-150 flux densities for J143213+350940 are $29.0\pm6.0$ and $46.3\pm4.6$\,mJy, respectively. If the LOFAR-150 flux density is correct, it would indicate that the spectrum is steep between 150 and 1400\,MHz, irrespective of whether the VLA-P or WENSS flux densities are correct. However, it is clear from Fig. \ref{fig:spectra} that the LOFAR-62 flux density upper limit shows that J143213+350940's spectrum either flattens towards lower frequencies or turns over irrespective of which of the four catalogues are correct at 153 and 325\,MHz.

Finally, we note that matching J143213+350940 to the AGES catalogue, we found a counterpart within $1.1$\,arcsec from our central position that has a photometric redshift of 1.02. The value agrees very well with those in Table \ref{tbl:targets}. 

\subsubsection{J143718+364549}
\label{subsec:J143718+364549}
\begin{figure}
  \includegraphics[width=\columnwidth]{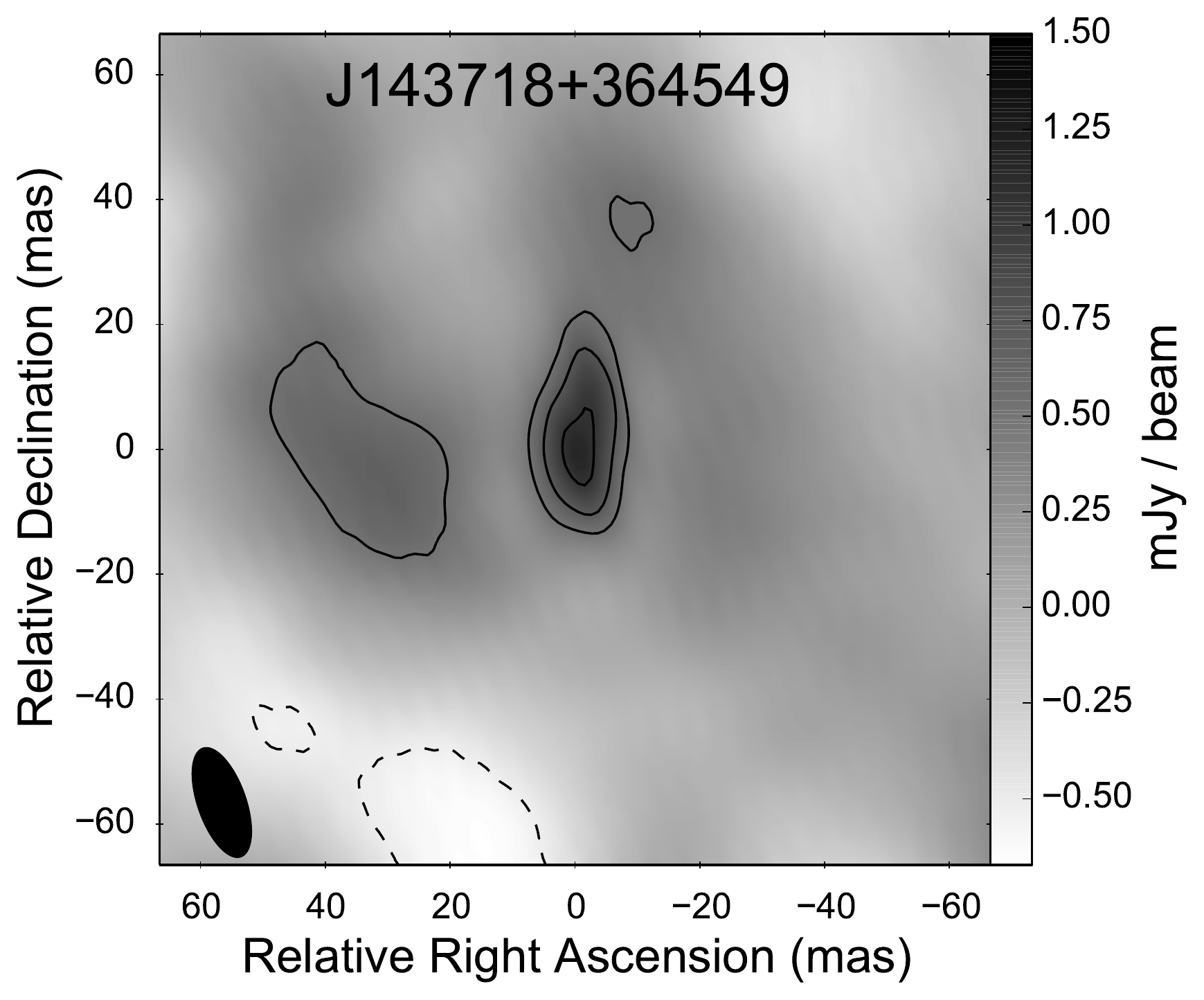}
  \caption{Naturally weighted EVN image of J143718+364549 without uv-tapering. J143718+364549 is significantly extended towards the North. The restoring beam is shown in the bottom left corner and has a size of $7.7\times18.4$\,mas at a major axis position angle of $19\fdg3$. The contours are drawn at $-3$ and 3 times the image noise, increasing in factors of $\sqrt 2$ thereafter.}
  \label{fig:J143718+364549}
\end{figure}

We selected J143718+364549 for observation based on its WENSS flux density. Unfortunately, because it is $2\fdg7$ from the phase center of the VLA-P image, we do not have a VLA-P flux density for it. The naturally weighted EVN image of J143718+364549 (Fig. \ref{fig:J143718+364549}) shows that the source is composed of a single resolved component that is significantly extended towards the North. To check if any flux density is resolved out in the naturally weighted image, we made an image in which we applied an uv-taper. The values derived from this image is reported in Table \ref{tbl:derived_info}. Tapering increased the recovered flux density from 2.9\,mJy to 6.1\,mJy, resulting in $68\pm9$\,per\,cent of the predicted flux density being recovered.

From J143718+364549's spectrum (Fig. \ref{fig:spectra}), it is clear that it either turns over or flattens towards lower frequencies. Finally, we point out that because we do not have a redshift for J143718+364549, the brightness temperature lower limit in Table \ref{tbl:derived_info} was derived using a redshift of zero.

\subsubsection{J144230+355735}
\label{subsec:J144230+355735}
Like J143718+364549, J144230+355735 lies outside the area imaged in the VLA-P image and was selected for observations with the EVN based on its WENSS flux density. From its spectrum (Fig. \ref{fig:spectra}), the source can either be turning over or flattening towards lower frequencies.

\begin{figure}
  \includegraphics[width=\columnwidth]{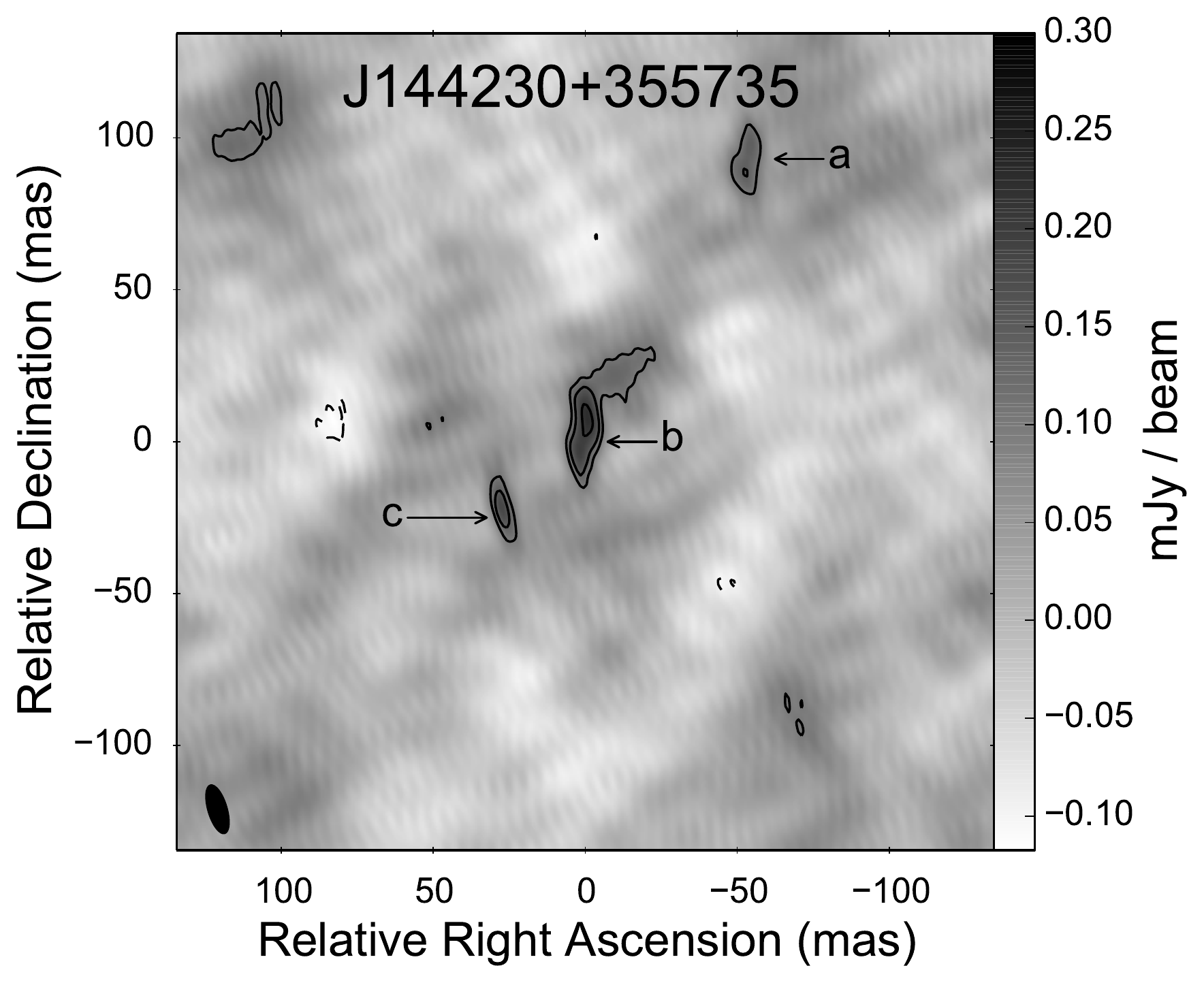}
  \caption{Naturally weighted EVN image of J144230+355735 without uv-tapering showing a CSO structure. The restoring beam is shown in the bottom left corner and has a size of $6.3\times16.7$\,mas at a major axis position angle of $16\fdg6$. The contours are drawn at $-3$ and 3 times the image noise, increasing in factors of $\sqrt 2$ thereafter.}
  \label{fig:J144230+355735}
\end{figure}

The EVN image of J144230+355735 (Fig. \ref{fig:J144230+355735}) is composed of three resolved components and has a CSO structure. J144230+355735's overall structure is that of a core with two lobes. From J144230+355735b a jet-like structure extends towards J144230+355735c.

To check if any of the flux density is resolved out in the naturally weighted image, we made an image in which we applied an uv-taper. The values derived from the uv-tapered image are reported for J144230+355735 in Table \ref{tbl:derived_info} while the values for J144230+355735a,b,c are derived from the non-uv-tapered image. Since the flux densities of J144230+355735b and J144230+355735c merge in the uv-tapered image, the component was fitted using a single elliptical Gaussian. In this image, J144230+355735a has a flux density of $1.07\pm0.13$\,mJy and a size of $46.7\pm3.4$\,mas. The combined component from J144230+355735b and J144230+355735c has a minor and major axis of $21.9\pm1.3$ and $64.0\pm3.7$\,mas, respectively, and a flux density of $1.26\pm0.11$\,mJy. Hence, after tapering, the total flux density of J144230+355735 increased by 0.6\,mJy resulting in $69\pm14$\,per\,cent of the predicted flux density being recovered. The missing flux density likely originates from low surface brightness emission along the jet axis between the central and outer components. It is however possible that at least some of the missing flux density is located beyond the outer components and that the jet is larger than what we measured. We note that while J144230+355735a is only detected at a $4.2\sigma$ level in Fig. \ref{fig:J144230+355735}, it is detected above $6\sigma$ in the uv-tapered image. Similarly the central position of J144230+355735c is encircled by a $6\sigma$ contour in the uv-tapered image.

The brightness temperature of the central component is consistent with the emission being non-thermal within its uncertainty. Furthermore, the brightness temperatures of all three components are lower limits as they were calculated using a redshift of zero since we could not determine a redshift for the source. Hence, if J144230+355735 is at a redshift above 0.45, the central component would clearly indicate AGN activity. Combining this with the structure of the source and the results in Section \ref{subsec:agn star}, we conclude that J144230+355735 is an AGN with a central core-jet structure and two lobes and/or hotspots.


\section{Discussion}
\label{sec:summary+conclude}

\subsection{What are the MPS sources?}
\label{subsec:what are mps}
From Sections \ref{subsec:agn star} and \ref{sec:Comments on Individual Sources}, we know that all the sources detected with the EVN are AGN. For the sources with redshifts, columns (11) and (12) in Table \ref{tbl:derived_info} show that they are GPS and CSS sources whose spectral turnovers have been redshifted to lower frequencies. Column (12) shows that J142850+345420 and J143718+364549, for which we do not have redshifts, are both smaller than $\sim1$\,kpc and are therefore either GPS or HFP sources. Hence, since all the sources that were detected with the EVN are either CSS, GPS or HFP sources, they are likely all young AGN \citep{o'dea1998,Murgia2002,Conway2002,2003PASA...20...19M,2009AN....330..120F}.

\begin{figure*}
  \includegraphics[width=\textwidth]{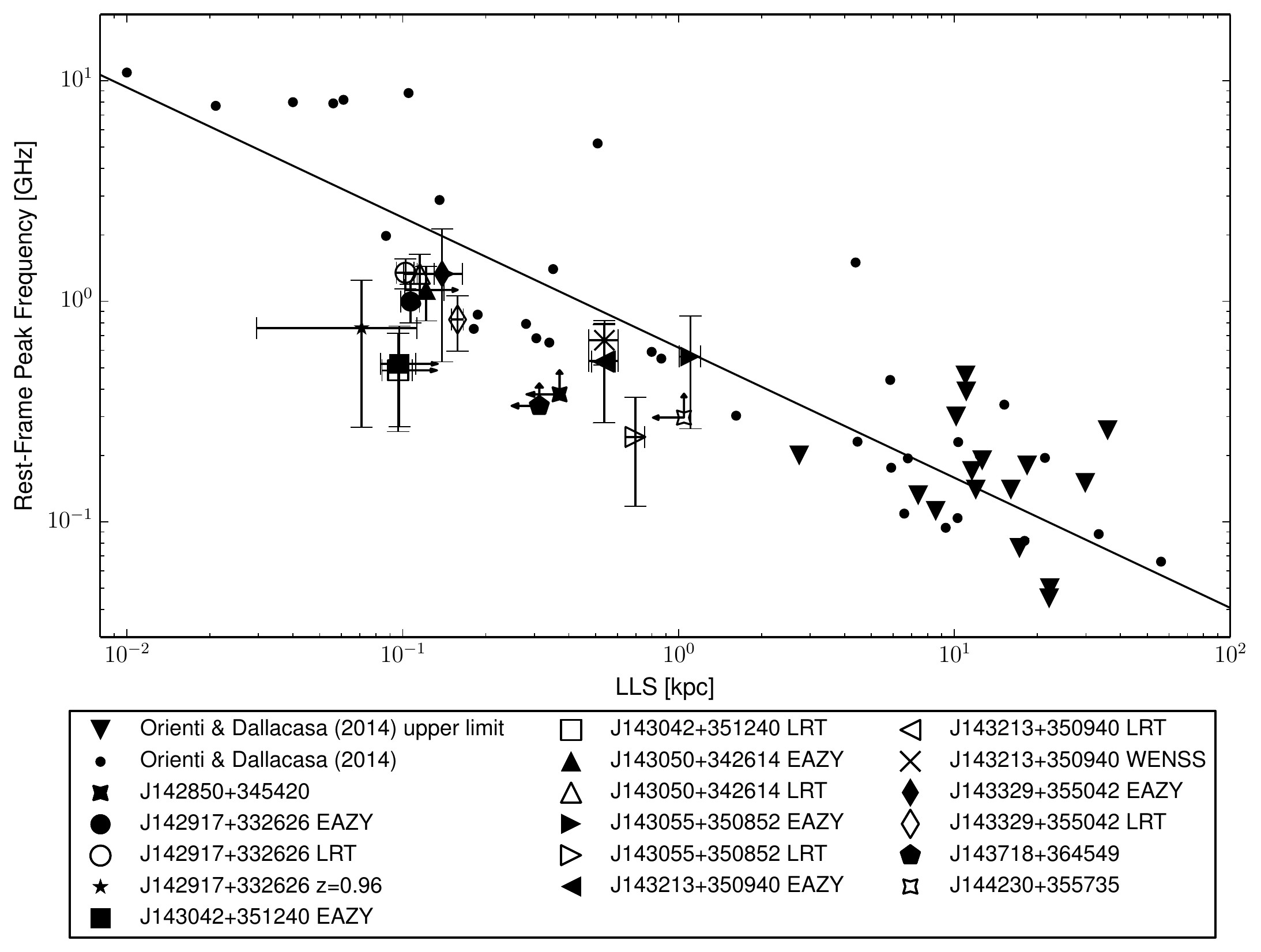}
  \caption{The figure used by \citet{Orienti2014} to derive the turnover frequency--linear size relation (shown as the solid line) with our sources included. The points and downward triangles show the sources from which \citet{Orienti2014} derived the relation. The downward triangles show sources for which the authors could only determine an upper limit for the turnover frequency. For each of our sources, the position is plotted using the redshifts from both the \textsc{eazy} and \textsc{lrt} codes. For J142850+345420, J143718+364549 and J144230+355735, for which we cannot calculate $\nu_{\mathrm r}$, we plotted $\nu_{\mathrm o}$ which is a lower limit for $\nu_{\mathrm r}$. Lower and upper limits on the LLS are indicated by arrows. A zoom-in of the region around our sources is shown in Fig. \ref{fig:trunover linear size zoom}.}
  \label{fig:trunover linear size}
\end{figure*}

\begin{figure}
  \includegraphics[width=\columnwidth]{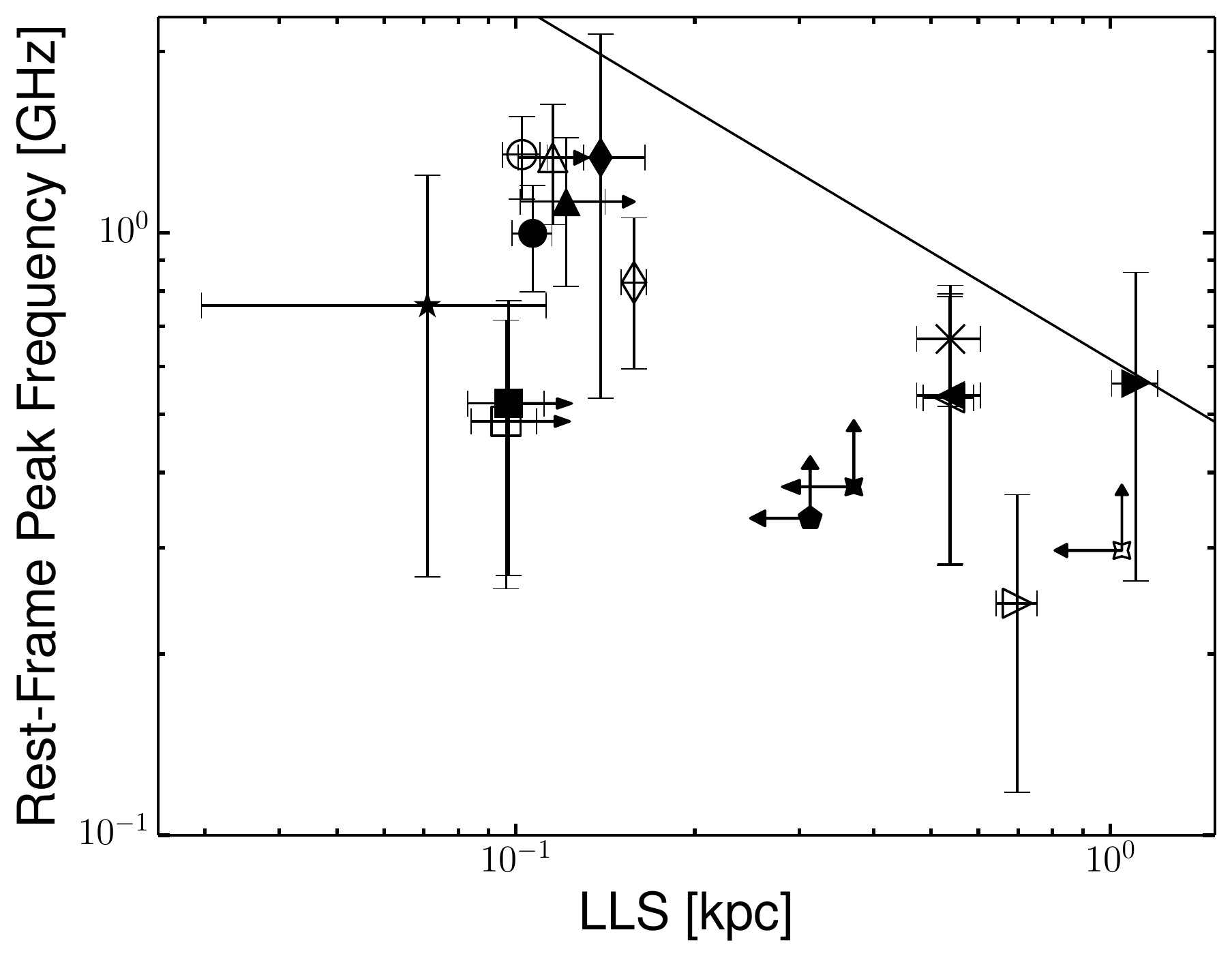}
  \caption{A zoomed-in image of the region around our sources in Fig. \ref{fig:trunover linear size} in which the sources from which \citet{Orienti2014} derived the relation was omitted for clarity. The figure legend is the same as that of Fig. \ref{fig:trunover linear size} and is not shown here for clarity. For J142850+345420, J143718+364549 and J144230+355735, for which we cannot calculate $\nu_{\mathrm r}$, we plotted $\nu_{\mathrm o}$ which is a lower limit for $\nu_{\mathrm r}$. Lower and upper limits are indicated by arrows.}
  \label{fig:trunover linear size zoom}
\end{figure}

In Fig. \ref{fig:trunover linear size}, we used the information in table 5 of \citet{Orienti2014} to generate the plot that the authors used to derive the turnover frequency--linear size relation. To determine the empirical relation, the authors compiled a list of CSOs for which the core components have been detected. The authors argue that this allows them to select genuine young radio galaxies and exclude other populations of sources such as blazars that could contaminate the sample. The resulting list of sources span redshifts between 0.08 and 2.93, linear sizes between 10\,pc and 56\,kpc and rest-frame turnover frequencies from 10.9\,GHz to below 45\,MHz. By minimizing the chi-squared error statistics, \citet{Orienti2014} found the relation
\begin{equation}
 \log_{10}\nu_{\mathrm r} = (-0.21\pm0.04) - (0.59\pm0.05)\log_{10} (\mathrm{LLS}),
 \label{eq:LLS turnover}
 \end{equation}
where $\nu_{\mathrm r}$ is in GHz and the LLS is in kpc. This relation agrees well with those found by both \citet{o'dea1998} and \citet{falcke2004}. Fig. \ref{fig:trunover linear size} shows the data used to derive the relation, with Fig. \ref{fig:trunover linear size zoom} zooming into the region around our sources.

To check if our sources lie within the scatter of points around the relation, we calculated the vertical distance between the relation and the sources used to derive it. From this we found that the median distance between the points and the line is 0.06 and has a $1\sigma$ deviation of 0.29. J143042+351240 EAZY and J143042+351240 LRT are $2.5\sigma$ and $2.7\sigma$ from the relation, with the remaining sources being within $2\sigma$ and J143050+342614 LRT, J143055+350852 EAZY, J143213+350940 EAZY, J143213+350940 LRT, J143213+350940 WENSS and J143329+355042 EAZY being within $1\sigma$. The maximum distance between the relation and the points used to derive it is 5.8\,GHz, which is more than the maximum distance between the relation and our sources. We note that since Equation \ref{eq:LLS turnover} has a constant slope, the vertical, horizontal and diagonal distances between each point and the relation are related to each other by constant factors. Hence, the number of $\sigma$ that each point is from the relation is independent of whether the vertical, horizontal or diagonal distance is used.

As discussed in Section \ref{subsec:J143213+350940}, the VLA-P and WENSS flux densities for J143213+350940 differ, resulting in a large uncertainty on the shape of its spectrum. Consequently, we plotted its position in Fig. \ref{fig:trunover linear size} and \ref{fig:trunover linear size zoom} for both the cases where $\nu_{\mathrm r}$ was derived using the VLA-P and WENSS flux densities. Since the \textsc{eazy} and \textsc{lrt} redshifts are so close to each other (Table \ref{tbl:targets}), the points are indistinguishable in the plots and we only plot the position of the \textsc{eazy} redshift point for the case where $\nu_{\mathrm r}$ was derived from the WENSS flux density. 

Since the redshift derived for J142917+332626 by \citet{Bouchefry2009} differs from that of the \textsc{eazy} and \textsc{lrt} values, we added an additional point (labeled as \lq J142917+332626 z=0.96') showing its position using the their redshift. J142917+332626 and J143055+350852 (for which the \textsc{eazy} and \textsc{lrt} redshift values differ significantly) illustrate the effect of redshift uncertainty on the position of the sources. Typically, an uncertainty in the redshift will have a larger effect on the rest-frame turnover frequency than on the linear size of the source, resulting in the sources primarily being displaced vertically. The reason for this is, for a source with a fixed linear size, its angular size does not simply decrease as a function of redshift (see the discussion on Column (12) of Table \ref{tbl:derived_info} in Section \ref{sec: results and discussion}), as is the case for its observed turnover frequency. Hence for sources around $z=1$ and above, an uncertainty in the redshift primarily has an affect on the rest-frame turnover frequency. However, for sources at $z<1$, a redshift uncertainty can have a significant effect on both $\nu_{\mathrm r}$ and the linear size.

Since we do not have redshifts for J142850+345420, J143718+364549 and J144230+355735, we used their observed turnover frequencies and size upper limits to plot their positions in Fig. \ref{fig:trunover linear size} and \ref{fig:trunover linear size zoom}. Since $\nu_{\mathrm r} = \nu_{\mathrm o}(1+z)$, $\nu_{\mathrm o}$ is a lower limit for $\nu_{\mathrm r}$. From the arguments in the previous paragraph, we expect that the true positions of the sources will primarily be vertical displaced compared to their current positions. As plotted, the sources are between 1.2 and $2.1\sigma$ from the relation.

Looking at Fig. \ref{fig:trunover linear size}, it is striking that all but two of our sources unambiguously lie below the correlation. There are several possible reasons for this, the first of which being that the sources were resolved out by the EVN, or that we missed components because they are to faint to be detected in the EVN images. In this case, the LLS will be underestimated and the sources should be towards the right of where they are plotted, closer to the relation. This is specifically the case for J143042+351240 and J143050+342614, for which, based on the percentage of the predicted flux density that was recovered from the EVN image, we concluded in Sections \ref{subsec:J143042+351240} and \ref{subsec:J143050+342614} that they are larger than their measured sizes. We also cannot rule out this possibility for the remainder of the detected sources (for which we recovered a larger percentage of the predicted flux density). For J143042+351240 and J143050+342614, the limits are indicated by arrows in Fig. \ref{fig:trunover linear size} and \ref{fig:trunover linear size zoom}. The second possibility why the sources lie below the correlation is selection effects. The first selection effect is that we can only select sources that have $\nu_{\mathrm o}\la500$\,MHz. Assuming that all of the sources that we can detect are at $z<10$, this excludes sources with $\nu_{\mathrm r}\ga5.5$\,GHz. The second selection effect is that the EVN's lack of short spacings prevents us from detecting extended emission on scales larger than $\sim35$\,mas. Hence, we cannot detect a source if it has a single, or multiple, components larger than $\sim35$\,mas, in which the flux density is uniformly distributed over the component or components. This is also the most likely explanation for why we did not detect J142904+354425 and J143024+352438. Since we cannot place an upper limit on the distance between two $\ga35$\,mas components, it is impossible to determine the true limitations imposed by this selection effect. Another selection effect is that the MPS sources were selected to be compact in FIRST (see Section 3.2 in \citet{coppejans2015}). However since this only sets the constraint that $\mathrm{LLS}\la54.5$\,kpc, this effect is irrelevant. In conclusion, we cannot therefore say whether the sources lying below the correlation is because of selection effects or not.

From the above it appears that the sources plotted in Fig. \ref{fig:trunover linear size} and \ref{fig:trunover linear size zoom} lie within the scatter of the points around the relation. We do note that the uncertainties on LLS and the peak frequency could influence this result. Since all the sources are AGN (Sections \ref{subsec:agn star} and \ref{sec:Comments on Individual Sources}), we therefore expect that at least some, if not all, of the sources will reveal themselves as CSOs if observed at high enough resolution. This can be said with certainty for, and is confirmed by, J143213+350940 and J144230+355735 which have CSO structures. We also point out again that, as discussed in Section \ref{subsec:variability}, it is very unlikely that any of the sources are blazars. 

As a final point, we note that the redshift range of the sources is similar to those of the sources from which \citet{Orienti2014} constructed the relation. Hence, if we assume that the sources are below the correlation purely because of their LLS's, and that they were not resolved out by the EVN, it is possible that they are less powerful than the sources in the relation and, therefore, get frustrated at smaller distances. Since AGN activity can be discontinuous \citep[e.g.][]{2000MNRAS.315..371S,2007MNRAS.378..581J}, it is also possible that their central engines have (temporarily) switched off. This would result in the core being undetected while the low surface density radio lobes continue to expand and fade. There are, however, two arguments against this. First, switched off AGN typically have spectra with $\alpha<-1.5$ \citep{1994A&A...285...27K,2015A&A...583A..89S}, which is steeper than the spectra of our sources. Secondly, the core accounts for a relatively small fraction of the total flux density \citep[e.g.][]{Orienti2014}, and therefore often remains undetected.

\subsection{Detection Fraction}
\label{subsec: detcetion frac}
In this section we compare our detection fraction of 0.818 to that of the mJy Imaging VLBA Exploration at 20\,cm \cite[mJIVE-20;][]{mJIVE} survey. mJIVE-20 is a VLBA survey to study the VLBI structure of FIRST sources. The survey has an angular resolution of 5\,mas, median noise of $0.178\mathrm{\,mJy\,beam^{-1}}$ and targets sources with FIRST peak intensities ($I_{\mathrm{FIRST}}$) between 1 and $200\mathrm{\,mJy\,beam^{-1}}$. To date, nearly 20\,000 sources have been observed of which 4336 were detected at or above a detection threshold of $6.75\sigma$, giving the survey an overall detection fraction of 0.217. Since our sources have $2.5<I_{\mathrm{FIRST}}<15.7$\,mJy\,beam$^{-1}$, we used the information in table 2 of \citet{mJIVE} to calculate that mJIVE-20 has a detection fraction of $0.151\pm0.007$ for sources with $2<I_{\mathrm{FIRST}}<16$\,mJy\,beam$^{-1}$. The fact that we have a higher detection fraction is in part be due to our lower detection threshold ($6\sigma$) and median noise ($0.095\mathrm{\,mJy\,beam^{-1}}$). To correct for this, we calculated that with mJIVE-20's median noise and detection threshold, our survey would detect 7 of our 11 sources. This means a detection fraction of 0.636, which is 3 times that of the mJIVE-20 survey as a whole and 4 times that achieved for the sources in our peak intensity range.

As discussed in Section \ref{subsec:what are mps}, MPS sources are some of the smallest AGN, which explains our high detection fraction in comparison to mJIVE-20. mJIVE-20 targets all sources with FIRST peak intensities above 1\,mJy\,beam$^{-1}$, while all our sources should be intrinsically small and thus good targets for VLBI.

\section{Summary and final points}
\label{sec:Summary}
In this paper, we presented high-resolution 1.7\,GHz EVN observations of 11 MPS sources. Of the 11 sources we detected nine with the EVN, recovering more than 50\,per\,cent of the predicted flux density for seven. Based on their radio luminosities, brightness temperatures and morphologies, we conclude that the radio emission of the sources detected with the EVN is from AGN activity (Sections \ref{subsec:agn star} and \ref{sec:Comments on Individual Sources}). In Section \ref{subsec:what are mps}, we used their peak frequencies and linear sizes to show that the MPS sources are a mixture of CSS, GPS and HFP sources whose spectral turnovers have been redshifted to lower frequencies. Hence, the MPS sources are likely all young AGN. We also argue that the detected sources are all likely CSOs. We do, however, emphasize that the uncertainties on the rest-frame turnover frequencies and the largest linear sizes could influence this conclusion. From their steep high-frequency spectra and brightness temperatures, we argue that none of the sources are blazars. We conclude that the radio emission from the two sources that were not detected with the EVN are likely from AGN activity, and that the reason for the non-detections is that they were resolved out by the EVN. Finally we conclude that our high detection fraction of 82\,per\,cent is due to the MPS sources being intrinsically small AGN, which makes them good targets for VLBI observations.

Because of small number statistics, we cannot say whether the sources at the highest redshifts are more compact than those at lower redshifts. However, all the detected sources, with the possible exceptions of J143042+351240 and J143050+342614, have linear and angular sizes smaller than 1.1\,kpc and 0.15\,arcsec, respectively. This shows that low-frequency colour-colour diagrams are an easy and efficient way of selecting small and likely young AGN. This also seems to indicate that when selecting a group of MPS sources that are compact on a scale of 5\,arcsec, most of the sources are compact on a scale below 0.5\,arcsec. Hence it is possible that to fulfill the criteria of selecting compact sources for the MPS method of searching for high-redshift AGN, all that is required is to select sources that are unresolved at 5\,arcsec.


\section*{Acknowledgements}

The authors wish to thank the anonymous referee, scientific editor and assistant editor for their suggestions and comments which helped to improve this paper.
The European VLBI Network is a joint facility of independent European, African, Asian, and North American radio astronomy institutes. Scientific results from data presented in this publication are derived from the following EVN project codes: EC053 (PI: R. Coppejans) and EV020 (PI: S. van Velzen).
The research leading to these results has received funding from the European Commission Seventh Framework Programme (FP/2007-2013) under grant agreement No 283393 (RadioNet3).
D.C. and S.F. thank the Hungarian Scientific Research Fund (OTKA NN110333) for their support.
C. M. and H. F. is funded by the ERC Synergy Grant BlackHoleCam: Imaging the Event Horizon of Black Holes (Grant 610058)
The authors would like to thank Cameron van Eck for his numerous helpful suggestions and insightful discussions.
We thank the staff of the GMRT that made these observations possible. The GMRT is run by the National Center for Radio Astrophysics of the Tata Institute of Fundamental Research.
Funding for SDSS-III has been provided by the Alfred P. Sloan Foundation, the Participating Institutions, the National Science Foundation, and the U.S. Department of Energy Office of Science. The SDSS-III web site is http://www.sdss3.org/.


\bibliographystyle{mnras.bst}
\bibliography{references.bib}

\label{lastpage}

\end{document}